\newcount\mgnf\newcount\tipi\newcount\tipoformule\newcount\greco

\tipi=2          
\tipoformule=0   


\global\newcount\numsec
\global\newcount\numfor
\global\newcount\numtheo
\global\advance\numtheo by 1

\def\senondefinito#1{\expandafter\ifx\csname#1\endcsname\relax}

\def\SIA #1,#2,#3 {\senondefinito{#1#2}%
\expandafter\xdef\csname #1#2\endcsname{#3}\else
\write16{???? ma #1,#2 e' gia' stato definito !!!!} \fi}

\def\etichetta(#1){(\veroparagrafo.\veraformula)%
\SIA e,#1,(\veroparagrafo.\veraformula) %
\global\advance\numfor by 1%
\write15{\string\FU (#1){\equ(#1)}}%
\write16{ EQ #1 ==> \equ(#1) }}

\def\letichetta(#1){\veroparagrafo.\verotheo
\SIA e,#1,{\veroparagrafo.\verotheo}
\global\advance\numtheo by 1
\write15{\string\FU (#1){\equ(#1)}}
\write16{ Sta \equ(#1) == #1 }}

\def\tetichetta(#1){\veroparagrafo.\veraformula 
\SIA e,#1,{(\veroparagrafo.\veraformula)}
\global\advance\numfor by 1
\write15{\string\FU (#1){\equ(#1)}}
\write16{ tag #1 ==> \equ(#1)}}

\def\FU(#1)#2{\SIA fu,#1,#2 }

\def\etichettaa(#1){(A\veroparagrafo.\veraformula)%
\SIA e,#1,(A\veroparagrafo.\veraformula) %
\global\advance\numfor by 1%
\write15{\string\FU (#1){\equ(#1)}}%
\write16{ EQ #1 ==> \equ(#1) }}

\def\BOZZA{
\def\alato(##1){%
 {\rlap{\kern-\hsize\kern-1.4truecm{$\scriptstyle##1$}}}}%
\def\aolado(##1){%
 {
{
 \rlap{\kern-1.4truecm{$\scriptstyle##1$}}}}} 
}

\def\alato(#1){}
\def\aolado(#1){}

\def\veroparagrafo{\number\numsec}
\def\veraformula{\number\numfor}
\def\verotheo{\number\numtheo}

\def\Eq(#1){\eqno{\etichetta(#1)\alato(#1)}}
\def\eq(#1){\etichetta(#1)\alato(#1)}
\def\leq(#1){\leqno{\aolado(#1)\etichetta(#1)}}
\def\teq(#1){\tag{\aolado(#1)\tetichetta(#1)\alato(#1)}}
\def\Eqa(#1){\eqno{\etichettaa(#1)\alato(#1)}}
\def\eqa(#1){\etichettaa(#1)\alato(#1)}
\def\eqv(#1){\senondefinito{fu#1}$\clubsuit$#1
\write16{#1 non e' (ancora) definito}%
\else\csname fu#1\endcsname\fi}
\def\equ(#1){\senondefinito{e#1}\eqv(#1)\else\csname e#1\endcsname\fi}

\def\Lemma(#1){\aolado(#1)Lemma \letichetta(#1)}%
\def\Theorem(#1){{\aolado(#1)Theorem \letichetta(#1)}}%
\def\Proposition(#1){\aolado(#1){Proposition \letichetta(#1)}}%
\def\Corollary(#1){{\aolado(#1)Corollary \letichetta(#1)}}%
\def\Remark(#1){{\noindent\aolado(#1){\bf Remark \letichetta(#1).}}}%
\def\Definition(#1){{\noindent\aolado(#1){\bf Definition 
\letichetta(#1)$\!\!$\hskip-1.6truemm}}}
\def\Example(#1){\aolado(#1) Example \letichetta(#1)$\!\!$\hskip-1.6truemm}

\def\include#1{
\openin13=#1.aux \ifeof13 \relax \else
\input #1.aux \closein13 \fi}

\openin14=\jobname.aux \ifeof14 \relax \else
\input \jobname.aux \closein14 \fi
\openout15=\jobname.aux

\let\EQ=\Eq


{\count255=\time\divide\count255 by 60 \xdef\hourmin{\number\count255}
        \multiply\count255 by-60\advance\count255 by\time
   \xdef\hourmin{\hourmin:\ifnum\count255<10 0\fi\the\count255}}

\def\oramin{\hourmin }

\def\data{\number\day/\ifcase\month\or january \or february \or march \or april
\or may \or june \or july \or august \or september
\or october \or november \or december \fi/\number\year;\ \oramin}

\newcount\pgn \pgn=1
\def\foglio{\number\numsec:\number\pgn
\global\advance\pgn by 1}
\def\foglioa{A\number\numsec:\number\pgn
\global\advance\pgn by 1}

\footline={\hss\tenrm\folio\hss}

\def\TIPIO{
\font\setterm=amr7 
\def \settepunti{\def\rm{\fam0\setterm}
\textfont0=\setterm   
\normalbaselineskip=9pt\normalbaselines\rm }\let\nota=\settepunti}

\def\TIPITOT{
\font\twelverm=cmr12
\font\twelvei=cmmi12
\font\twelvesy=cmsy10 scaled\magstep1
\font\twelveex=cmex10 scaled\magstep1
\font\twelveit=cmti12
\font\twelvett=cmtt12
\font\twelvebf=cmbx12
\font\twelvesl=cmsl12
\font\ninerm=cmr9
\font\ninesy=cmsy9
\font\eightrm=cmr8
\font\eighti=cmmi8
\font\eightsy=cmsy8
\font\eightbf=cmbx8
\font\eighttt=cmtt8
\font\eightsl=cmsl8
\font\eightit=cmti8
\font\sixrm=cmr6
\font\sixbf=cmbx6
\font\sixi=cmmi6
\font\sixsy=cmsy6
\font\twelvetruecmr=cmr10 scaled\magstep1
\font\twelvetruecmsy=cmsy10 scaled\magstep1
\font\tentruecmr=cmr10
\font\tentruecmsy=cmsy10
\font\eighttruecmr=cmr8
\font\eighttruecmsy=cmsy8
\font\seventruecmr=cmr7
\font\seventruecmsy=cmsy7
\font\sixtruecmr=cmr6
\font\sixtruecmsy=cmsy6
\font\fivetruecmr=cmr5
\font\fivetruecmsy=cmsy5
\textfont\truecmr=\tentruecmr
\scriptfont\truecmr=\seventruecmr
\scriptscriptfont\truecmr=\fivetruecmr
\textfont\truecmsy=\tentruecmsy
\scriptfont\truecmsy=\seventruecmsy
\scriptscriptfont\truecmr=\fivetruecmr
\scriptscriptfont\truecmsy=\fivetruecmsy
\def \eightpoint{\def\rm{\fam0\eightrm}
\textfont0=\eightrm \scriptfont0=\sixrm \scriptscriptfont0=\fiverm
\textfont1=\eighti \scriptfont1=\sixi   \scriptscriptfont1=\fivei
\textfont2=\eightsy \scriptfont2=\sixsy   \scriptscriptfont2=\fivesy
\textfont3=\tenex \scriptfont3=\tenex   \scriptscriptfont3=\tenex
\textfont\itfam=\eightit  \def\it{\fam\itfam\eightit}%
\textfont\slfam=\eightsl  \def\sl{\fam\slfam\eightsl}%
\textfont\ttfam=\eighttt  \def\tt{\fam\ttfam\eighttt}%
\textfont\bffam=\eightbf  \scriptfont\bffam=\sixbf
\scriptscriptfont\bffam=\fivebf  \def\bf{\fam\bffam\eightbf}%
\tt \ttglue=.5em plus.25em minus.15em
\setbox\strutbox=\hbox{\vrule height7pt depth2pt width0pt}%
\normalbaselineskip=9pt
\let\sc=\sixrm  \let\big=\eightbig  \normalbaselines\rm
\textfont\truecmr=\eighttruecmr
\scriptfont\truecmr=\sixtruecmr
\scriptscriptfont\truecmr=\fivetruecmr
\textfont\truecmsy=\eighttruecmsy
\scriptfont\truecmsy=\sixtruecmsy }\let\nota=\eightpoint}

\newfam\msbfam   
\newfam\truecmr  
\newfam\truecmsy 
\newskip\ttglue
\ifnum\tipi=0\TIPIO \else\ifnum\tipi=1 \TIPI\else \TIPITOT\fi\fi

\def\E{{I\kern-.25em{E}}}
\def\N{{I\kern-.25em{N}}}
\def\M{{I\kern-.25em{M}}}
\def\R{{I\kern-.25em{R}}}
\def\Z{{Z\kern-.425em{Z}}}
\def\1{{1\kern-.25em\hbox{\rm I}}}
\def\eu{{1\kern-.25em\hbox{\sm I}}}

\def\C{{I\kern-.64em{C}}}
\def\P{{I\kern-.25em{P}}}
\def\eop{{ \vrule height7pt width7pt depth0pt}\par\bigskip}



\def\chap #1#2{\line{\ch #1\hfill}\numsec=#2\numfor=1}

\def\sqr#1#2{{\vcenter{\vbox{\hrule height.#2pt
     \hbox{\vrule width.#2pt height#1pt \kern#1pt
   \vrule width.#2pt}\hrule height.#2pt}}}}
\def\qed{ $\mathchoice\sqr64\sqr64\sqr{2.1}3\sqr{1.5}3$} 


\newcount\foot
\foot=1
\def\note#1{\footnote{${}^{\number\foot}$}{\ftn #1}\advance\foot by 1}
\def\tag #1{\eqno{\hbox{\rm(#1)}}}
\def\frac#1#2{{#1\over #2}}

\def\text#1{\quad{\hbox{#1}}\quad}

\def\thanks{\noindent{\bf Aknowledgements: }}



\font\ch=cmbx12
\font\ftn=cmr8

\font\it=cmti10
\font\bf=cmbx10
\font\sm=cmr7

%
\catcode`\X=12\catcode`\@=11
\def\n@wcount{\alloc@0\count\countdef\insc@unt}
\def\n@wwrite{\alloc@7\write\chardef\sixt@@n}
\def\n@wread{\alloc@6\read\chardef\sixt@@n}
\def\crossrefs#1{\ifx\alltgs#1\let\tr@ce=\alltgs\else\def\tr@ce{#1,}\fi
   \n@wwrite\cit@tionsout\openout\cit@tionsout=\jobname.cit 
   \write\cit@tionsout{\tr@ce}\expandafter\setfl@gs\tr@ce,}
\def\setfl@gs#1,{\def\@{#1}\ifx\@\empty\let\next=\relax
   \else\let\next=\setfl@gs\expandafter\xdef
   \csname#1tr@cetrue\endcsname{}\fi\next}
\newcount\sectno\sectno=0\newcount\subsectno\subsectno=0\def\r@s@t{\relax}
\def\resetall{\global\advance\sectno by 1\subsectno=0
  \gdef\firstpart{\number\sectno}\r@s@t}
\def\resetsub{\global\advance\subsectno by 1
   \gdef\firstpart{\number\sectno.\number\subsectno}\r@s@t}
\def\v@idline{\par}\def\firstpart{\number\sectno}
\def\l@c@l#1X{\firstpart.#1}\def\gl@b@l#1X{#1}\def\t@d@l#1X{{}}
\def\m@ketag#1#2{\expandafter\n@wcount\csname#2tagno\endcsname
     \csname#2tagno\endcsname=0\let\tail=\alltgs\xdef\alltgs{\tail#2,}%
  \ifx#1\l@c@l\let\tail=\r@s@t\xdef\r@s@t{\csname#2tagno\endcsname=0\tail}\fi
   \expandafter\gdef\csname#2cite\endcsname##1{\expandafter
     \ifx\csname#2tag##1\endcsname\relax?\else{\rm\csname#2tag##1\endcsname}\fi
    \expandafter\ifx\csname#2tr@cetrue\endcsname\relax\else
     \write\cit@tionsout{#2tag ##1 cited on page \folio.}\fi}%
   \expandafter\gdef\csname#2page\endcsname##1{\expandafter
     \ifx\csname#2page##1\endcsname\relax?\else\csname#2page##1\endcsname\fi
     \expandafter\ifx\csname#2tr@cetrue\endcsname\relax\else
     \write\cit@tionsout{#2tag ##1 cited on page \folio.}\fi}%
   \expandafter\gdef\csname#2tag\endcsname##1{\global\advance
     \csname#2tagno\endcsname by 1%
   \expandafter\ifx\csname#2check##1\endcsname\relax\else%
\fi
   \expandafter\xdef\csname#2check##1\endcsname{}%
   \expandafter\xdef\csname#2tag##1\endcsname
     {#1\number\csname#2tagno\endcsnameX}%
   \write\t@gsout{#2tag ##1 assigned number \csname#2tag##1\endcsname\space
      on page \number\count0.}%
   \csname#2tag##1\endcsname}}%
\def\m@kecs #1tag #2 assigned number #3 on page #4.%
   {\expandafter\gdef\csname#1tag#2\endcsname{#3}
   \expandafter\gdef\csname#1page#2\endcsname{#4}}
\def\re@der{\ifeof\t@gsin\let\next=\relax\else
    \read\t@gsin to\t@gline\ifx\t@gline\v@idline\else
    \expandafter\m@kecs \t@gline\fi\let \next=\re@der\fi\next}
\def\t@gs#1{\def\alltgs{}\m@ketag#1e\m@ketag#1s\m@ketag\t@d@l p
    \m@ketag\gl@b@l r \n@wread\t@gsin\openin\t@gsin=\jobname.tgs \re@der
    \closein\t@gsin\n@wwrite\t@gsout\openout\t@gsout=\jobname.tgs }
\outer\def\localtags{\t@gs\l@c@l}
\outer\def\globaltags{\t@gs\gl@b@l}
\outer\def\newlocaltag#1{\m@ketag\l@c@l{#1}}
\outer\def\newglobaltag#1{\m@ketag\gl@b@l{#1}}

\def\t@gsoff#1,{\def\@{#1}\ifx\@\empty\let\next=\relax\else\let\next=\t@gsoff
   \expandafter\gdef\csname#1cite\endcsname{\relax}
   \expandafter\gdef\csname#1page\endcsname##1{?}
   \expandafter\gdef\csname#1tag\endcsname{\relax}\fi\next}
\def\verbatimtags{\let\ift@gs=\iffalse\ifx\alltgs\relax\else
   \expandafter\t@gsoff\alltgs,\fi}
\catcode`\X=11 \catcode`\@=\active
\localtags
%
\setbox200\hbox{$\scriptscriptstyle \data $}
\global\newcount\numpunt
\magnification=\magstephalf
\hoffset=0.cm
\baselineskip=14pt  
\parindent=12pt
\lineskip=4pt\lineskiplimit=0.1pt
\parskip=0.1pt plus1pt

\hyphenation{small}
 
\count0=1


\def\qed{\lower 2 pt \vbox{\hrule width 8 pt height 8 pt depth 0 pt}}
\def\boxin#1{\lower 3.5 pt \vbox{\hrule \hbox{\strut \vrule{} #1 \vrule} 
\hrule}}

 2

\def\sne{S^{N-1}(\sqrt{E})}
\def\hh{{\cal H}_{N,E}}

\def\ncht{\left(\matrix{N\cr 2\cr}\right)}
\def\nmcht{\left(\matrix{N-1\cr 2\cr}\right)}

\catcode`\@=11 

\centerline{\bf Determination of the Spectral Gap for Kac's Master Equation}
\centerline{\bf and Related Stochastic Evolutions} 

\vskip 1.5truecm
 \centerline{E. A. Carlen\footnote{$^\star$}{\eightpoint carlen@math.gatech.edu,
 loss@math.gatech.edu, Work 
 partially supported by
U.S. N.S.F. grant DMS 00-70589 }\quad
 M. C. Carvalho\footnote{$^{\star\star}$}{\eightpoint mcarvalh@math.gatech.edu,  On leave from Departamento do Mathematic, Fac. di Ciences de Lisboa,
 partially supported by PRAXIS XXI and TMR ERB-FMRX CT97 0157}\quad
M. Loss{$^\star$}} 
\vfootnote{}{\baselineskip = 12pt{\eightpoint\copyright 2001 by the authors.
Reproduction of this
article, in its entirety, by any means is permitted for non-commercial
purposes.}}
\bigskip 
\centerline{School of Mathematics}
\centerline{  Georgia Institute of Technology}
\centerline{ Atlanta, GA, 30332 U.S.A.}

\bigskip
\bigskip {\baselineskip = 12pt\narrower{\noindent {\bf Abstract }\/} 
We present a method for bounding, and in some cases computing, the spectral gap
for systems of many particles evolving under the influence of a random 
collision mechanism. 
In particular,
the method yields the exact spectral gap in a model due to Mark Kac of energy conserving 
collisions with one dimensional velocities. 
It is also sufficiently robust to  provide qualitatively sharp
bounds also in the case of more physically realistic
momentum and energy conserving collisions in three dimensions, as well as
a range of related models.} 


\vfill\break
\chap {Introduction }1

We derive sharp bounds on the rate of relaxation to equilibrium for two models of random
collisions connected with the Boltzmann equation, as well as several other stochastic
evolutions of a related type.  In fact, there is a fairly broad class of models to which the
methods used here may be applied.  The starting point is a model due to Mark Kac [\rcite{K}]
of random energy preserving ``molecular collisions'', and its analysis provides the pattern
for the analysis of all of the models discussed here, including a more physically realistic
model of random energy and momentum conserving collisions.  However, since the features of
the Kac model have motivated the method of analysis presented here, we begin by introducing
it.

The Kac model represents a system of $N$ particles in one dimension evolving under a 
random collision mechanism. 
It is assumed that the spatial distribution of the particles
is uniform, so that the state of the system is given by 
specifying the $N$ velocities
$v_1,v_2,\dots,v_N$. 
The random collision mechanism under which the state evolves is that
at random times $T_j$, a ``pair collision'' takes place in such a way that
the total energy
$$E = \sum_{k=i}^N v_k^2\Eq(A1)$$
is conserved. Since only a pair of one dimensional velocities is involved in each collision,
there are just two degrees of freedom active, and if the collisions were to conserve both 
energy and momentum, the only possible non--trivial result of a collision would be an
exchange of the two velocities. Since Kac sought a model in which the distribution 
of the
velocities would equilibriate over the energy surface specified by (1.1), he dropped the
requirement of momentum conservation, and retained only energy conservation.

With  energy conservation being the only constraint on a pair collision,
the kinematically possible ``post--collisional'' 
velocities
when particles $i$ and $j$ collide, 
$v^*_i$ and $v^*_j$, are of the form
$$v^*_i(\theta) = v_i\cos(\theta) + v_j\sin(\theta)\qquad{\rm and}\qquad
v^*_j(\theta) = -v_i\sin(\theta) + v_j\cos(\theta)\ \Eq(A2)$$
where, of course, $v_i$ and $v_j$ are the pre--collisional velocities, and $\theta \in (-\pi,\pi]$.

To specify the evolution, consider it first in discrete time, 
collision by collision. Let
$${\vec v}(k) =\left(v_1(k),v_2(k), \dots,v_N(k)\right)\Eq(A3)$$
denote the state of the system just after the $k$th collision. Evidently, ${\vec v}(k)$
is a random variable with values in $S^{N-1}(\sqrt{E})$, the sphere in $R^N$
of radius $\sqrt{E}$, where $E$ is the energy. Let $\phi$ be 
any continuous function on $S^{N-1}(\sqrt{E})$. We will specify the
collision mechanism by giving a formula for computing the conditional
expectation of $\phi({\vec v}(k+1))$ given ${\vec v}(k)$, which defines the
one--step Markov transition operator $Q$ through
$$ Q\phi({\vec v}) = {\rm E}\{\phi({\vec v}(k+1))\ |\ {\vec v}(k) = {\vec v}\ \}\ .
\Eq(A4)$$

In the collision process to be modeled, the pair $\{i,j\}$, $i<j$, 
of molecules that collide is to be selected uniformly at random. Then
the velocities $v_i$ and $v_j$ are updated by choosing an 
angle $\theta$, and letting (1.2) define the post--collisional velocities.
Let $\rho(\theta)$ be a probability density on the circle; i.e,
$$\int_{-\pi}^{\pi}\rho(\theta){\rm d}\theta = 1\ ,\Eq(A5)$$
and take $\rho$ to be the probability density for the outcome that the collision results in 
post--collisional velocities $v^*_i(\theta)$ and $v^*_j(\theta)$ as in \eqv(A2).

Then for any  
function $\phi$ on $\R^n$, and hence by restriction on $S^{N-1}(\sqrt{E})$,
$${\rm E}\{\phi({\vec v}(k))\ |\ {\vec v}(k-1) = {\vec v}\ \} = 
{\ncht}^{-1}\sum_{i<j}^N\int_{-\pi}^{\pi}\rho(\theta)
f(v_1,v_2,\dots,v^*_i(\theta),\dots,v^*_j(\theta),\dots,v_n){\rm d}\theta\ ,$$
and thus
$$Q\phi ({\vec v}) = 
{\ncht}^{-1}\sum_{i<j}^N\int_{-\pi}^{\pi}\rho(\theta)
\phi (v_1,v_2,\dots,v^*_i(\theta),\dots,v^*_j(\theta),\dots,v_n)
{\rm d}\theta\ .\Eq(A6)$$
This expression can be clarified if for each $i<j$ we let $R_{i,j}(\theta)$ denote the 
rotation in $\R^N$ that induces a clockwise rotation in the $v_i,v_j$ plane through an angle $\theta$,
and fixes the orthogonal complement of this plane. Then $R_{i,j}(\theta)\vec v$ is 
the post--collisional velocity vector corresponding to the 
pre--collisional velocity vector $\vec v$, and \eqv(A6) can be rewritten as
$$Q\phi ({\vec v}) = 
{\ncht}^{-1}\sum_{i<j}^N\int_{-\pi}^{\pi}\rho(\theta)
\phi (R_{i,j}(\theta)\vec v)
{\rm d}\theta\ .\Eq(A6a)$$

Let $\hh$ denote the Hilbert space of square integrable functions $f$
on the sphere 
$S^{N-1}(\sqrt{E})$ 
equipped with the normalized uniform measure ${\rm d}\mu_N$. 
Let $\langle\cdot,\cdot\rangle$ and $\|\cdot\|$ 
denote the inner product
and norm on $\hh$.
We now require that $\rho(\theta) = \rho(-\theta)$ so that $Q$  is  
self adjoint on $\hh$. We also require that $\rho$ be continuous and strictly positive at $\theta= 0$,
which is a convenient condition ensuring that $Q$ is ergodic.

It is clear from \eqv(A6a) that $Q$ is an average over isometries on
$\hh$, and hence is 
a contraction. 
Moreover,
$\|Qf\|_2 = \|f\|_2$
if and only if $f$ is constant by our ergodicity assumptions on $\rho$.

Because $Q$ is self adjoint, 
it updates the probability density $f_k$ for ${\vec v}$ as well.  
Indeed,  for any
test function $\phi$,
$$\eqalign{
\int_{S^{N-1}(\sqrt{E})}\phi({\vec v})f_{k+1}({\vec v}){\rm d}\mu_N &=  
{\rm E}\phi({\vec v}(k+1))\cr 
&= {\rm E}\left({\rm E}\{\phi({\vec v}(k+1)\ |\ {\vec v}(k)\ \}\right)
={\rm E} Q\phi({\vec v}(k))\cr
&=\int_{S^{N-1}(\sqrt{E})}\phi({\vec v})Qf_{k}({\vec v}){\rm d}\mu_N\cr}$$
which of course means that $Qf_k = f_{k+1}$.

One passes to a continuous time description by letting the waiting times between collisions become
continuously distributed random variables. To obtain a Markov process,
the distribution of these waiting times must be memoryless, and hence
exponential.
Therefore, fix some parameter $\tau_N>0$, and define the Markovian semigroup
$G_t$, $t>0$, by
$$G_t f = e^{-(t/ \tau_N)} \sum_{k=0}^\infty {(t/\tau_N)^k\over k!}Q^k f =
e^{(t/\tau_N)(Q- I)}f\ ,$$
which gives the evolution of the probability density for ${\vec v}$, continuously in 
the time $t$.

It remains to specify the dependence of $\tau_N$ on $N$. Let $T^{(N)}$ denote the waiting time
between collisions in the $N$--particle model. 
Suppose that the waiting time for any given particle to undergo 
a collision is  independent of $N$, which corresponds roughly 
to adjusting the size of the container
with $N$ so that the particle density remains constant. Suppose also that
these waiting times  are all independent of one another,
which {\it should} be reasonable for a gas of many particles. 
(See Kac [\rcite{K}] for further discussion.)
Then we would have
$Pr\{ T^{(N)}_j > t\} = Pr\{ T^{(1)}_j > t\}^N$,
or 
$e^{-t\tau_N} = e^{- NT/\tau_1}$.
Changing the time scale, we put $\tau_1 =1$ and hence $\tau_N = 1/N$.
Therefore, the semigroup is given by
$G_t  = e^{tN(Q-I)}$. For any initial probability density $f_0$, $f(\vec v,t) = G_tf_0(\vec v)$
solves {\it Kac's Master Equation}
$${\partial\over \partial t}f(\vec v,t) = N(Q - I)f(\vec v,t)\ ,\Eq(kacmast)$$
which is the evolution equation for the model in  so far as we are concerned with the probability density
$f(\vec v,t)$ for the velocities at time $t$, and not the velocities $\vec v(t)$ themselves, 
which are random variables.

Because of the ergodicity, if $f_0$ is any initial probability density for the
process, it is clear that
$\lim_{t\to\infty}G_tf_0 = 1$.
The question is how fast this relaxation to the invariant density $1$ occurs.
To quantify this,
define
$$\lambda_N = \sup\left\{{1\over N}\langle f, Q f\rangle\ \biggl|
\ \|f\|_2=1\ ,\langle f, 1\rangle=0\ \right\}\ .\Eq(lamndef)$$ 
If this quantity is
less than one, there is a ``gap'' in the spectrum of $Q$, and hence
the {\it spectral gap} of $Q$ is defined to be
$1 - \lambda_N$.
Clearly then, the spectral gap for $N(Q-I)$ is 
$$\Delta_N = N(1 - \lambda_N)\ .\Eq(delenndef)$$ 
This  quantity is of interest in quantifying the rate of relaxation of
$G_tf_0$ to $1$ since for 
any square integrable initial probability
density $f_0$,
$$\|G_t(f_0-1)\|_2 \le e^{-t\Delta_N}\|f_0-1\|_2\ .$$

Mark Kac, who introduced this operator and process [\rcite{K}] in 1956,  
observed that for each fixed 
$\ell$, the subspace of spherical harmonics of degree $\ell$ in $\sne$ is 
an invariant subspace under $Q$. (This is especially clear from \eqv(A6a) since if
$\phi$ is in such a subspace, then so is $\phi\circ R_{i,j}(\theta)$
for each pair $i<j$ and each angle $\theta$). 
Since each of these subspaces is finite dimensional,
$Q$ has a pure point spectrum.
He remarks that it is not even evident that $\Delta_N > 0$
for all $N$, much less that there is a lower bound independent of $N$. 
(As Diaconis and Saloff--Coste noted in [\rcite{DSC}], $Q$ is not compact.) He nonetheless 
conjectured that
$$\liminf_{N\to \infty}\Delta_N = C > 0\ .\Eq(kaccon)$$

Kac's conjecture in this form, for the special case
$\rho = 1/2\pi$ considered explicitly by Kac,  was recently proved by Janvresse [\rcite{J}]
using Yau's martingale method [\rcite{Y1}], [\rcite{Y2}]. Her proof 
gives no information on the value of $C$.
One result we prove here is that in the case $\rho = 1/2\pi$,
$$\Delta_N = {1\over 2}{N+2\over N-1}\ ,\Eq(skaccon1)$$
and hence
$$\liminf_{N\to \infty}\Delta_N = {1\over 2}\ .\Eq(skaccon)$$

The result \eqv(skaccon1)
has also been obtained by Maslin in unpublished work, using entirely different methods. Some
account of Maslin's results can be found in a paper [\rcite{DSC}] by Diaconis and Saloff--Coste
in which it is shown that $\Delta_N \ge C/N^2$ for the Kac model as well as a natural 
generalization of it
in which the sphere $S^N$ is replaced by the special orthogonal group $SO(N)$.
Our method gives exact results in this case too, as we shall see.

Maslin's approach was   based on the representation theory of the group $SO(N)$, 
and does not seem to extend to
more general cases, such as a non-uniform density $\rho(\theta)$, or to momentum 
conserving collisions.
According to his former thesis advisor, Persi Diaconis,  
this is one reason it was never published.
We will comment further on the relation of our paper to previous work, especially [\rcite{J}],
 [\rcite{Y1}], [\rcite{Y2}] and [\rcite{DSC}], in section 3 
where we carry out
our analysis of the Kac model, and in section 6 where we analyze the $SO(N)$ 
variant of the Kac model.

Kac did not explicitly conjecture \eqv(skaccon), only \eqv(kaccon), though he discussed 
motivations for his conjecture that
do suggest \eqv(skaccon).
In particular, he was motivated by a connection between the many--particle evolution described 
by the Master equation \eqv(kacmast), 
and 
a model Boltzmann equation, and he did rigorously establish the following connection: 
For each integer $k$, $1 \le k \le N$, let
$\pi_k$ be the $k$th coordinate projection on
$\sne$; i.e., 
$$\pi_k(v_1,v_2,\dots,v_N) = v_k\ .\Eq(AApid)$$
Then define the orthogonal projections $P_k$ through
$$\langle P_k f\circ\pi_k, g\circ\pi_k\rangle = 
\langle  f, g\circ\pi_k\rangle$$
for all $f$ in $\hh$, and all continuous 
bounded functions $g$ on $[-\sqrt{E},\sqrt{E}]$. That is, $P_k$ is the 
orthogonal projection onto the subspace in ${\cal H}_{N,E}$ consisting of
functions of the form $g\circ\pi_k$. In probabilistic language, $P_kf$ is 
the conditional expectation 
of $f$ given $v_k$; i.e.,
$$P_kf(v) = {\rm E}\{\ f \ |\ v_k = v\ \}\ ,\Eq(AAApid)$$ 
and when $f$ is a probability density on $\sne$, $P_kf$ is its $k$th 
single particle marginal. Kac showed that
if a sequence of initial densities $f_0^{(N)}$ on $\sne$ satisfies a 
certain symmetry and independence
property that he called ``molecular chaos'',  and if furthermore 
$$g(v) = \lim_{N\to \infty}P_1f_0(v)$$
exists in $L^1(\R)$, then so does  $g(v,t) = \lim_{N\to \infty}P_1\left(G_tf_0(v)\right)$, and
$g(v,t)$ satisfies the {\it Kac Equation}
$${\partial\over \partial t}g(v,t)  = 
2\int_{-\pi}^{\pi}\left(\int_{\R}\left[g(v^*(\theta),t)g(w^*(\theta),t) -
g(v,t)g(w,t)\right]{\rm d}w\right)\rho(\theta){\rm d}\theta\ .\Eq(kacequ)$$
The fact that there is a quadratic non--linearity on the right is due to 
the fact that the underlying
many particle dynamics is generated by pair collisions. 
The factor of 2 on the right hand side comes from the 2
in the the normalization factor $2/N(N-1)$ in the definition of $Q$, \eqv(A6). 
The $N$ is absorbed by the factor of
$N$ in $N(Q- I)$, the generator of $G_t$, and the $N-1$ is absorbed by 
summing over all the $N-1$ particles with which the
first particle can collide. 

Kac's limit theorem provides a direct link between the linear but many particle Master equation \eqv(kacmast)
and the one variable but non--linear Kac  equation \eqv(kacequ). Kac's proposal was that one should 
be able to obtain quantitative results about the behavior of the Master equation, and from these, 
deduce quantitative results on the Kac equation \eqv(kacequ). Specifically, he was concerned with following this route
to results on the rate of relaxation to equilibrium for solutions of \eqv(kacequ).

It is easy to see that for any $\beta>0$, 
$$m_\beta(v) = \sqrt{\beta\over 2\pi}e^{-\beta v^2/2}\ \Eq(maxwell)$$
is a steady state solution of the Kac equation \eqv(kacequ). (In the kinetic
theory context, the Gaussian density in \eqv(maxwell) is known 
as the {\it Maxwellian} density with temperature $1/\beta$).
Indeed, 
as is well known, $m_\beta$ is the limit of the
single particle marginal on $S^{N-1}(\sqrt{N/\beta})$ as $N$ tends to infinity. 
Kac wanted to show that for any reasonable
initial data
$g(v)$, the Kac equation had a solution $g(v,t)$ with $\lim_{t\to\infty}g(v,t) = 
m_\beta(v)$ where
$\int_{\R}v^2g(v){\rm d}v = 1/\beta$. Indeed, he wanted to show that this 
convergence took place {\it exponentially
fast}, and he boldly conjectured that one could prove this exponential 
convergence for the Master equation from whence \eqv(kacequ) came. At
the time Kac wrote his paper, very little was known about the non--linear
Boltzmann equation, Carleman's 1933 paper [\rcite{TC}] being one of the
few mathematical studies. Given the difficulties inherent in dealing
directly with the non--linear equation, his  suggested approach via
the Master eqaution was well motivated, though unfortunately he  did
not suceed himself in obtaining quantitative relaxation estimates by
this route, and other workers choose to directly investigate the non--linear equation.

Evidence for  the conjectured exponential convergence 
came from linearizing the Kac equation about the steady state solutions $m_\beta$. The resulting 
generator of the linearized Kac equation can be written 
in terms of averages of Mehler kernels, as shown in [\rcite{M}],
and so all of the eigenfunctions are Hermite polynomials (as Kac had observed in  section
9 of [\rcite{K}]).
The eigenvalue corresponding to the $n$th degree Hermite polynomial, $n\ge 1$,  
is then readily  worked out  to be (see [\rcite{M}]):
$$2\int_{-\pi}^{\pi}\left(\sin^n(\theta) + \cos^n(\theta) - 1
\right)\rho(\theta){\rm d}\theta\ .\Eq(kaceqev)$$
The eigenvalue is zero for $n=2$, corresponding to conservation of energy.
As we have indicated, Kac actually only
considered the special case in which $\rho$ was uniform; i.e., 
$\rho(\theta) = 1/2\pi$. In this case, 
the eigenvalues are
$-1$ for
$n$ odd, and are monotonically decreasing toward $-1$ for $n$ even. 
Thus, the eigenvalue corresponding  to the to fourth
degree Hermite polynomial determines the spectral gap for the 
linearization of \eqv(kacequ) in this case. {\it The fact that this gap is 
$1/2$ is consistent with \eqv(skaccon), and bears out Kac's intuition that there is a close
quantitative connection between his Master equation \eqv(kacmast), and the Kac equation \eqv(kacequ).}

In fact, as we shall see, in the
case considered by Kac and some other cases as well,
$\lambda_N$ is an eigenvalue of $Q$ of multiplicity one, and
$Qf_N(\vec v) = \lambda_N f_N(\vec v)$
for
$$f_N(v_1, \dots, v_N) = \sum_{j=1}^N\left(v_j^4 - \langle 1,v_j^4\rangle\right)\ .\Eq(evform1)$$
If  $E = N$ and $g_N$ is defined by 
$P_1(f_N) = g_N\circ\pi_N$,
then 
$$\lim_{N\to\infty}g_N(v) = m_1(v) h_{(4)}(v)\ ,\Eq(Az8z8)$$
where $h_{(4)}$ is the fourth degree Hermite polynomial 
for the standard unit variance Gaussian measure on $\R$.
(This is fairly evident, but will be fully evident in view of the 
formula for $P_1$ given in section 2.)
Thus, the correspondence between the spectral gaps extends to a correspondence between the eigenfunctions too.

McKean [\rcite{M}] and Gruenbaum [\rcite{G1}], [\rcite{G2}] have further investigated these
issues. In particular, McKean conjectured  that reasonable solutions of \eqv(kacequ)
should relax to the Gaussian stationary solutions of the same energy in 
$L^1$ at the exponential rate $e^{-t/2}$ corresponding to the 
spectral gap in the linearized equation. He proved this for nice initial data
but with exponential rate $e^{-tc}$ where $c$ is explicit constant, but about an order of magnitude
smaller than $1/2$. Later, in [\rcite{CGT}] 
this result was established with the almost the sharp rate, i.e., $e^{-[(1/2)-\epsilon]t}$
for nice initial data. See the papers for precise statements,
but note that all of this is in the case $\rho =1/2\pi$.
(The results  are stated differently in [\rcite{M}] and  [\rcite{CGT}], which
use a different time scale so that the
factor of $2$ in \eqv(kacequ) is absent). 

If one expects that the linearized version of \eqv(kacequ) is a good guide to 
the behavior solutions of \eqv(kacequ), one might guess  that \eqv(kaceqev)
provides a good guide to the relaxation properties of solutions
of \eqv(kaceqev). This would suggest that in the case in which $\rho$
is uniform, the slowest mode of relaxation corresponds to initial
data of the form $m_1(v)(1 + \epsilon h_{(4)}(v)))$ for small $\epsilon$.

If one further believed that the non-linear Kac equation \eqv(kacequ)
is a good guide to behavior of solutions of Kac's Master equation,
then one might guess that the slowest mode of relaxation for the Master
equation is a symmetric fourth degree ploynomial, at least for uniform $\rho$. 
Such a line of
reasoning suggests \eqv(evform1) as a candidate for the slowest mode
of relaxation for Kac's Master equation. This turn out to be correct,
as we have indicated, and this shows how well--constructed the
Kac model is: A great deal of information is washed out and lost
whenever one passes from the $N$--particle distribtion function $f(\vec v)$
to its single particle marginal  distribution $g(v)$. In general, there would be
no reason to expect that the slowest mode of decay for the Master 
equation would not be lost in passing to the marginal.

Indeed, it is easy to see that
$f_N$ is in fact an eigenfunction of $Q$. 
We shall see that  for many choices of $\rho$, $f_N$ is the optimizer in
\eqv(lamndef). 
This correspondence between Kac's Master equation \eqv(kacmast) and the linearized  
version of the Kac equation \eqv(kacequ) is a full vindication of Kac's conjectures. It also shows that his model is free of extraneous detail at the microscopic level; 
what happens at the microscopic level described by the Master equation
is what happens at the level described by \eqv(kacequ).

We conclude the introduction by briefly stating our results for the Kac model itself, 
and then describing the structure of the paper. The key result in our analysis of the Kac model is
the following theorem which reduces the variational problem \eqv(lamndef) to a much simpler, purely
geometric, one dimensional problem:

\bigskip
\noindent{\bf Theorem 1.1} {\it For all $N\ge 3$, 
$$\Delta_N \ge (1- \kappa_N)\Delta_{N-1}\Eq(recA)$$
where
$$\kappa_N = \sup\left\{{\langle g\circ\pi_1, g\circ\pi_2\rangle
\over \|g\circ\pi_1\|_2}\ \biggl|\ g\in {\cal C}([-1,1])\ ,
\  \langle g\circ\pi_1, 1\rangle=0\ \right\}\Eq(varA)$$
}
\medskip

Notice first of all that $g$ is a function of a single variable -- in contrast to \eqv(lamndef),
\eqv(varA) is a one--dimensional variational problem. Also notice that \eqv(varA) doesn't involve $\rho$,
or otherwise  directly refer to $Q$.

The bound in Theorem 1.1 implies that
$\liminf_{N\to\infty}\Delta_N \ge 
\prod_{J=3}^\infty(1 - \kappa_j)\Delta_2$.
Since the necessary and sufficient condition for the infinite product to be non-zero 
is that
$$\sum_{j=3}^\infty \kappa_j < \infty\ ,\Eq(Az9z9)$$
proving the Kac conjecture is reduced to the problem of proving the summability of
$\kappa_j$ and the strict positivity of $\Delta_2$. 

The second part is easy, since
for  two particles, $Q$ is an operator on functions on $S^1$.
Indeed
$$ \langle f,Qf\rangle = 
{1\over 2\pi}\int_{-\pi}^\pi \int_{-\pi}^\pi f(\psi)f(\psi-\theta)\rho(\theta){\rm d}\theta{\rm d}\psi \ ,
\Eq(AQ2p)$$ 
and writing this in terms of Fourier series leads to
$$\lambda_2 = \sup_{k\ne 0}
\left\{\int_{-\pi}^{\pi}\rho(\theta)\cos(k\theta){\rm d}\theta\right\}\ .
\Eq(AQ2y)$$
By the Riemann--Lebesgue lemma, $\lambda_2 <1$, and hence $\Delta_2 = 2(1-\lambda_2) > 0$. 

As for the summability of $\kappa_N$, note from \eqv(varA) that
$\kappa_N$ is a measure of the dependence of the 
coordinate functions on the sphere. With $E=N$,
the marginal distribution of $(v_1,v_2)$ induced by $\mu_N$ is
$${|S^{N-3}|\over N |S^{N-1}|}
\left(1 - {v_1^2 + v_2^2\over N}\right)^{(N-4)/2}{\rm d}v_1{\rm d}v_2\ .$$
As $N$ tends to infinity, this tends to
$${1\over 2\pi}e^{-(v_1^2+v_2^2)/2}{\rm d}v_1{\rm d}v_2\ ,$$
and under this limiting measure, the two coordinate functions $v_1$ and $v_2$ are independent. 
Hence
for any admissible trial function $g$ in \eqv(varA),
$$\eqalign{
&\lim_{N\to\infty}\langle g\circ\pi_1, g\circ\pi_2\rangle
 =\cr
&{1\over 2\pi}\int_{\R^2}g(v_1)g(v_2)e^{-(v_1^2+v_2^2)/2}{\rm d}v_1{\rm d}v_2 =\cr
&{1\over \sqrt{2\pi}}\int_{\R}g(v_1)e^{-v_1^2/2}{\rm d}v_1
 {1\over \sqrt{2\pi}}\int_{\R}g(v_2)e^{-v_1^2/2}{\rm d}v_2 = \cr
&\lim_{N\to\infty}\langle g\circ\pi_1, 1\rangle \langle g\circ\pi_2, 1\rangle
= 0\cr}\Eq(limAAA)$$
which implies that $\lim_{N\to\infty}\kappa_N = 0$, without, however,  showing how fast. 

In fact, it is not hard to compute $\kappa_N$:
\bigskip
\noindent{\bf Theorem 1.2} {\it For all $N\ge 3$,
$$\kappa_N = {3\over N^2-1}\ .$$}
\medskip
Since this is summable, \eqv(Az9z9) holds, and so the Kac conjecture is proved.
But Theorem 1.2 tell us much more than just \eqv(Az9z9). One can exactly solve
the recurrence relation in Theorem 1 with $\kappa_N = 3/(N^2-1)$. 
As we shall see, this leads to:

\bigskip
\noindent{\bf Theorem 1.3} {\it For all $N\ge 2$,
$$\Delta_N \ge {1 - \lambda_2\over 2}{N+2\over N-1}\ .\Eq(bnd1A)$$
Moreover, this result is sharp for the case considered by Kac; i.e., constant
density $\rho$, in which case $\lambda_2 =0$,
and more generally whenever
$$\int_{-\pi}^{\pi}\rho(\theta)\cos(k\theta){\rm d}\theta \le 
\int_{-\pi}^{\pi}\rho(\theta)\cos(4\theta){\rm d}\theta\Eq(4cond)$$
for all $k\ne 0$. In all of these cases, $\lambda_N$ has multiplicity one,
and the corresponding eigenfunction is
$$\sum_{k=1}^N\left(v_k^4 - \langle v_k^4,1\rangle\right)\ .\Eq(4evev)$$
 }\bigskip

Notice that $\Delta_N$ does not depend on $E$. 
This is easy to see directly from the fact that $Q$
commutes with the unitary change of scale that relates $\hh$ and 
${\cal H}_{N,E'}$ for two different values,
$E$ and $E'$, of the energy. We have kept it 
present in the discussion until now on account of the relation between
the uniform probability measure on $\sne$, and the unit Gauss measure on the 
real line, or on $\R^2$ as in \eqv(limAAA).
Having said what we wish to say about this, it will be simplest to 
henceforth set $E=1$, and to delete it from our
notation.

The division of our results on the original Kac model into 
Theorems 1.1, 1.2 and 1.3 of course reflects
the steps in the method by which they are obtained. However, 
it also reflects a point of physical relevance, namely
{\it  that $\kappa_N$ is completely independent of 
$\rho(\theta)$.} The complicated details of the collision mechanism do not enter into
$\kappa_N$. Rather, they enter our estimate for $\Delta_N$ {\it only} 
through the the value of the {\it two 
particle} gap $\Delta_2 = 2(1 - \lambda_2)$. Once this is computed, 
there is a {\it purely geometric}
relation between the values of the gap for different values of $N$. The fact
that there should be such a simple and purely geometric relation between
the values of the gap for different values of $N$ is a very interesting feature 
of the Kac model which expresses the strong sense in which it is a binary collision model.

The paper is organized as follows: In Section 2 we identify the  general features 
of the Kac model
that enable us to prove Theorem 1.1. We then introduce the notion of 
a {\it Kac system}, which embodies
these features, and prove the results that lead to analogs of Theorem 1.1 
for general Kac systems.
This provides a convenient framework for the analysis of a number of models, 
as we illustrate in the
next four sections. Section 3 is devoted to the Kac model itself, and 
contains the proofs of Theorems 1.1,
1.2 and 1.3. Section 4 is devoted to the analysis of the master equation for 
physical, three dimensional, 
momentum and energy conserving Boltzmann collisions.  Sections 5 is devoted to 
a shuffling model that has been studied in full detail
by Diaconis and Shahshahani [\rcite{DS}]. We include this here because it 
can be viewed 
as the Kac model with momentum conservation, and is very simple. 
(We hasten to add that
Diaconis and Shahshahani do much more 
for this model than compute the spectral gap). Then in Section 6 we treat 
another generalization 
of the Kac model, this time 
in the direction of greater complexity: The $SO(N)$ model of Maslin, 
Diaconis and Saloff--Coste [\rcite{DSC}]. 
Finally, in Section 7 we show that the quartic eigenfunction \eqv(evform1) is indeed 
the gap eigenfunction for a wide range of nonuniform densities $\rho$ that
violate \eqv(4cond).

\bigskip

\chap {2: General Features}2

\bigskip

The Kac model introduced in the previous section has the following general features that are shared 
by all of the models discussed here:
\medskip
\noindent{\bf Feature 1:} {\it For each $N>1$ there is measure space 
$(X_N,{\cal S}_N,\mu_N)$, with $\mu_N$ a probability measure, on which there 
is a measure preserving action of $\Pi_N$,
the symmetric group on $N$ letters. We denote
$${\cal H}_N = L^2(X_N,\mu_N)\ .\Eq(chd)$$}
\medskip
We think of $X_N$ as the ``$N$ particle phase space'' or
``$N$ particle state space'', and the action of $\Pi_N$
as representing ``exchange of particles''. In the Kac model, $X_N$ is $S^{N-1}$,
${\cal S}_N$ is the Borel field, and $\mu_N$ is the 
rotation invariant probability measure on $S^{N-1}= X_N$. A permutation $\sigma\in \Pi_N$ acts on $X_N$
through 
$$\sigma(v_1,v_2,\dots,v_N) = (v_{\sigma(1)},v_{\sigma(2)},\dots,v_{\sigma(N)})\ .$$
\medskip

\noindent{\bf Feature 2:} {\it There is another measure space 
$(Y_N,{\cal T}_N,\nu_N)$ and there are measureable maps $\pi_j:X_N\rightarrow Y_N$
for $j =1,2,\dots,N$ such that for all $\sigma\in \Pi_N$, and each $j$,
$$\pi_j\circ\sigma = \pi_{\sigma(j)}\ .\Eq(perminv)$$
Moreover, for each $j$, and all $A \in {\cal T}_N$,
$$\nu_N(A) = \mu_N(\pi_j^{-1}(A))\ .\Eq(push)$$
We denote
$${\cal K}_N = L^2(Y_N,\nu_N)\ .\Eq(chd)$$}
\medskip

We think of $\pi_j(x)$ as giving the ``state of the $j$th particle when the $N$ particle system is in 
state $x$''. For example, in the Kac model, we take
$$\pi_j(v_1,v_2,\dots,v_N) = v_j \in [-1,1]\Eq(kpidef)$$
and thus we take $Y_N = [-1,1]$. In this case, $Y_N$ does not depend on $N$, and it may seem strange to allow 
the single particle state space itself to depend on $N$. However, the methods we use here permit this generality, 
and some of the examples considered here require it.

Notice that once $Y_N$ and the $\pi_j$ are given, $\nu_N$ is specified through \eqv(push). 
In the Kac model we therefore have
$$\nu_N(v) = {|S^{N-2}|\over |S^{N-1}|}(1-v^2)^{(N-3)/2}{\rm d}v\ .\Eq(knudef)$$

\medskip

\noindent{\bf Feature 3:} {\it For each $N\ge 3$ and each $j = 1,2\dots,N$, there is a map
$$\phi_j: \left(X_{N-1}\times Y_N\right)\rightarrow X_N\Eq(triv)$$
so that 
$$\pi_j(\phi_j(x,y)) = y\Eq(back1)$$
for all $j=1,\dots,N$ and all $(x,y)\in X_{N-1}\times Y_N$. Moreover, $\phi_j$ has the property that
for all $A \in {\cal S}_N$,
$$\left[\mu_{N-1}\otimes\nu_N\right](\phi_j^{-1}(A)) = \mu_N(A)\ ,\Eq(tens)$$
or equivalently, for all bounded measurable functions $f$ on $X_N$,
$$\int_{X_N}f{\rm d}\mu_N = \int_{Y_N}\left[\int_{X_{N-1}}f(\phi_j(x,y)){\rm d}\mu_{N-1}(x)\right]
{\rm d}\nu_N(y)\ .\Eq(tens2)$$
}
\medskip

In the Kac model case, for any $\tilde v\in X_{N-1}=S^{N-2}$ and any $v\in Y_N =[-1,1]$ we put
$$\phi_N(\tilde v,v) = (\sqrt{1-v^2}\tilde v,v)\ ,\Eq(kembed)$$
and \eqv(tens) is easily verified.

So far, none of the features we have considered involve the dynamics.  
That is, the first three features
are purely kinematical. The fourth feature brings in the  Markov 
transition operator $Q$. We do not make the dependence of $Q$ on $N$ explicit in our notation, 
since this will always be clear from the context.

\medskip

\noindent{\bf Feature 4:} {\it For each $N\ge 2$, there is 
a selfadjoint and positivity preserving  operator $Q$ on ${\cal H}_N$ such that
$Q1 =1$. These operators are related to one another by the following:
For each $N\ge 3$, each $j = 1,2,\dots,N$, and each square integrable
function $f$ on $X_N$,
$$\langle f,Qf\rangle_{{\cal H}_N} = {1\over N}\sum_{j=1}^N\int_{Y_N}\left(
\langle f_{j,y},Qf_{j,y}\rangle_{{\cal H}_{N-1}}
\right){\rm d}\nu_{N}(y)\ \Eq(ave)$$
where for each $j$ and each $y\in Y_N$,
$$ f_{j,y}(\cdot) = f(\phi_j(\cdot,y))\ .\Eq(pass)$$   }
\medskip
\noindent{It} is easily verified that the Kac model posseses this feature. 

\medskip
\noindent{\bf Definition:} A {\it Kac System} is a system of probability spaces 
$(X_N,{\cal S}_N,\mu_N)$ and $(Y_N,{\cal T}_N,\nu_N)$ for $N\in \N$, $N\ge 2$, together with, for each $N$,
maps $\pi_j$ and $\phi_j$, $j = 1,2,\dots,N$,  a measure preserving action of $\Pi_N$ on 
$(X_N,{\cal S}_N,\mu_N)$, and a Markov transition operator $Q$ on ${\cal H} =L^2(X_N,\mu_N)$,
related to one another in such a way that they possess all of the
properties specified in features 1 through 4 above.
\medskip

In analyzing the spectral gaps of the operators $Q$ in Kac systems, 
certain other operators related to conditional expectations will play a central role, as indicated
in the previous section. Suppose that
$(X_N,{\cal S}_N,\mu_N)$, $(Y_N,{\cal T}_N,\nu_N)$, $\pi_j$, and $\phi_j$, are defined and related
as specified above. For each $j=1,2,\dots , N$, let $P_j$ be the orthogonal projection onto the subspace
of ${\cal H}_N$ consisting of functions of the form $g\circ \pi_j$ for some $g\in {\cal K}_N$.
Then, with $y = \pi_j(x)$ and $f_{j,y}$ given by \eqv(pass),
$$P_jf(x) = g(\pi_j(x)) \qquad{\rm where}\qquad g(y) = \int_{X_{N-1}}f_{j,y}(\tilde x){\rm d}\mu_{N-1}(\tilde x)\
.\Eq(pjdef)$$

In terms of these projections, define
$$P = {1\over N}\sum_{j=1}^N P_j\Eq(pdef)$$
which is clearly a positive contraction on ${\cal H}_N$. Define a contraction $K$ on
${\cal K}_N$ by
$$(Kg)\circ\pi_N = P_N\left(g\circ \pi_{N-1}\right)\ .\Eq(kdef)$$
Note that $Kg(y)$ is the conditional expectation of $g\circ\pi_2$ given that $\pi_1=y$. 
That is,
$$Kg(y) = {\rm E}\{g\circ\pi_N \ |\ \pi_{N-1} = y\ \}\ .\Eq(BQ2y)$$
In concrete examples, it
is easy to deduce an explicit formula for $K$ from \eqv(kdef) and \eqv(pjdef)
or directly from \eqv(BQ2y).
By the permuation symmetry,
$$P_i\left(g\circ \pi_j\right) = (Kg)\circ\pi_i\qquad{\rm for\ all}\qquad i\ne j\ .\Eq(also)$$
Combining \eqv(pass), \eqv(pjdef) and \eqv(kdef), we obtain
$$Kg(y) = \int_{X_{N-1}}g\left(\pi_{N-1}(\phi_{N}(\tilde x,y))\right){\rm d}\mu_{N-1}(\tilde x)\Eq(kker)$$
which provides an explicit form for the operator $K$. For example, in the case of the Kac model we obtain
$$\eqalign{Kg(v)) &= \int_{X_{N-1}}g(\sqrt{1- v^2}w_{N-1}){\rm d}\mu_{N-1}(w)\cr
&= \int_{-1}^1g(\sqrt{1- v^2}w){\rm d}\nu_{N-1}(w)\cr
&= {|S^{N-3}|\over |S^{N-2}|}\int_{-1}^1g(\sqrt{1- v^2}w)(1-w^2)^{(N-4)/2}{\rm d}w\cr}\Eq(kack)$$
from \eqv(knudef) and \eqv(kker). 
\medskip
\noindent{\bf Theorem 2.1} {\it Given any Kac system, let $P$ and $K$ be defined by
\eqv(pdef) and \eqv(kdef). Define $\mu_N$, $\kappa_N$ and $\beta_N$ by
$$\mu_N = \sup\{\ \langle f,Pf\rangle_{{\cal H}_N}\ |\  \|f\|_{{\cal H}_N} =1\quad{\rm and}\quad
\langle 1,f\rangle_{{\cal H}_N}= 0\ \}\Eq(mudef)$$
$$\kappa_N = \sup\{\ \langle g,Kg\rangle_{{\cal K}_N}\ |\  \|g\|_{{\cal K}_N} =1\quad{\rm and}\quad
\langle 1,g\rangle_{{\cal K}_N}= 0\ \}\Eq(kadef)$$
$$\beta_N = {1\over N-1}|\inf\{\ \langle g,Kg\rangle_{{\cal K}_N}\ |\  \|g\|_{{\cal K}_N} =1 \}|\ .
\qquad\qquad\qquad\qquad\ \phantom{1}\Eq(bedef)$$
Suppose, moreover, that the operator $P$ has pure point spectrum. Then, either $\mu_N = 0$ or
$$\mu_N = \max\left\{  {1\over N}(1 + (N-1)\kappa_N)\ ,\ {1\over N}(1 + (N-1)\beta_N)\ \right\}
\ .\Eq(muform)$$
In case $\kappa_N > \beta_N$, then the multiplicity of $\kappa_N$ as an eigenvalue of $K$
coincides with the multiplicity of $\mu_N$ as an eigenvalue of $P$, and the map
$$h \mapsto \left({1 \over N(1 +(N-1)\kappa_N)}\right)^{1/2}\sum_{j=1}^N h\circ \pi_j\Eq(nop4)$$
is an isometry from the $\kappa_N$--eigenspace of $K$ in ${\cal K}_N$ to the  
$\mu_N$--eigenspace of $P$ in ${\cal H}_N$.}
 
\noindent{\bf Proof:} 
Suppose that $f$ is an eigenfunction of $P$ with eigenvalue $\nu$.
Since $P$ commutes with permutations, 
we may assume that either $f$ is invariant under permutations, or that there is some transposition,
which we may as well take to be $\sigma_{1,2}$, such that $f\circ\sigma_{1,2} = -f$. 
We will treat these two cases separately.

First suppose that $f$ is symmetric.
Then for some $h\in {\cal K}_N$ independent of $k$, $P_kf = h\circ \pi_k$, and so
$$\nu f = Pf = {1\over N}\sum_{k=1}^N P_kf = {1\over N}\sum_{k=1}^N h\circ \pi_k\ .\Eq(WS1)$$
where $\nu$ is the eigenvalue, and $h$ is some function on $\R$.
Applying $P_1$ to both sides of \eqv(WS1) yields
$$ \nu P_1 f = {1\over N}\sum_{k=1}^N P_1 (h\circ \pi_k) \ ,$$
which can be easily simplified to
$$ \nu h \circ \pi_1 = {1\over N} (h \circ \pi_1 + (N-1) Kh \circ \pi_1) \ .$$
If $h$ is identically zero, i.e., if $P_jf=0$ for all $j=1, \dots , N$, the eigenvalue $\nu$ 
must necessarily be zero by \eqv(WS1).
Otherwise, $h$ is an eigenfunction of $K$ with eigenvalue $\tilde \nu$ so that 
$$\nu = {1\over N}(1 + (N-1)\tilde \nu)\ .\Eq(SKP)$$
Note that $\nu$ may vanish without $h$ being the zero function. In this case $h$ must 
be an eigenfunction of $K$ with eigenvalue $-1/(N-1)$. In any case, $\nu=0$ or otherwise equation
\eqv(SKP) must hold.

We next consider the case in which 
$$f\circ\sigma_{1,2} = -f\ .$$
Note that
$$P_k(f\circ\sigma_{1,2}) = P_kf=0$$
whenever $k$ is different from both $1$ and $2$. It follows that
$${1\over N}\sum_{k=1}^NP_kf = {1\over N}\left(P_1f + P_2f\right)\ .$$

Again, applying $P_1$ to both sides of the equation and keeping in mind that when
$P_1f = h \circ \pi_1$ then $P_2 f = -h \circ \pi_2$ we get
$$\nu h \circ \pi_1 = {1 \over N} (h \circ \pi_1 - Kh \circ \pi_1) \ .$$
In case $h$ vanishes identically, the eigenvalue $\nu$ must vanish also, otherwise
$h$ is an eigenfunction of $K$ with eigenvalue $\tilde \nu$ such that
$$\nu = {1- \tilde\nu \over N}\ .\Eq(NSKP)$$
Since by the definition \eqv(bedef), $-(N-1)\beta_N$ is the most negative eigenvalue of $K$, 
the determination of $\mu_N$ now follows from \eqv(SKP) and \eqv(NSKP). 

For the last part, observe from \eqv(SKP) and \eqv(NSKP) that when $\kappa_N > \beta_N\ge 0$,
$$\mu_N = {1\over N}(1 + (N-1) \kappa_N)\ ,\Eq(SKP3)$$
and, by the argument leading to \eqv(SKP), any eigenfunction of $P$ with eigenvalue $\mu_N$
is in the range of the map in \eqv(nop4), and a simple computation verifies the isometry property.
\eop
\medskip
\noindent{\bf Theorem 2.2} {\it Given any Kac system, let $P$ and $\mu_N$ be defined by
\eqv(pdef) and \eqv(mudef). Define $\lambda_N$ by
$$\lambda_N = 
\sup\{
\langle f, Q f\rangle_{{\cal H}_N}\ |\ 
\|f\|_{{\cal H}_N}=1\ ,\ \langle f, 1\rangle_{{\cal H}_N} =0\ \}\ .\Eq(lamdef)$$
Then
$$\lambda_N \le \left( \lambda_{N-1} + (1 - \lambda_{N-1})\mu_N\right)\ .\Eq(basrec)$$
Moreover, there is equality in \eqv(basrec) if and only if the suprema in \eqv(lamdef)
and \eqv(mudef) are attained at a common function $f_N$.}
\bigskip
\noindent{\bf Proof:} We start from \eqv(ave), taking any function $f\in {\cal H}_N$ 
satisfying the conditions imposed in \eqv(lamdef).
$$\eqalign{\langle f,Qf\rangle_{{\cal H}_N} &= {1\over N}\sum_{j=1}^N\int_{Y_N}
\langle f_{j,y},Qf_{j,y}\rangle_{{\cal H}_{N-1}}
{\rm d}\nu_{N}(y)\cr
&={1\over N}\sum_{j=1}^N\int_{Y_N}
\langle \left[f_{j,y} - P_jf(y)\right]+P_jf(y),Q
\left(\left[f_{j,y} - P_jf(y)\right]+P_jf(y)\right)\rangle_{{\cal H}_{N-1}}
{\rm d}\nu_{N}(y)\cr
&={1\over N}\sum_{j=1}^N\int_{Y_N}\
\langle [f_{j,y} - P_jf(y)],Q
[f_{j,y} - P_jf(y)]\rangle_{{\cal H}_{N-1}}
{\rm d}\nu_{N}(y)\cr
&+ {1\over N}\sum_{j=1}^N
\int_{Y_N}|P_jf(y)|^2{\rm d}\nu_{N}(y)
\ ,\cr}$$
Since each $P_jf(y)$ is constant on $X_{N-1}$ and so on ${\cal H}_{N-1}$, $QP_jf(y) = P_jf(y)$, and
$$\langle \left[f_{j,y} - P_jf(y)\right],P_jf(y)\rangle_{{\cal H}_{N-1}} = 0 \ .$$ 
But
$${1\over N}\sum_{j=1}^N
\int_{Y_N}|P_jf(y)|^2{\rm d}\nu_{N}(y) = \langle f, P f\rangle_{{\cal H}_N}\ ,$$
and hence
$$\eqalign{\langle f,Qf\rangle_{{\cal H}_N} 
&={1\over N}\sum_{j=1}^N\int_{Y_N}\
\langle [f_{j,y} - P_jf(y)],Q
[f_{j,y} - P_jf(y)]\rangle_{{\cal H}_{N-1}}
{\rm d}\nu_{N}(y)\cr
&+ \langle f, Pf \rangle_{{\cal H}_N}
\ ,\cr}\Eq(mid)$$
Now since $\langle[f_{j,y} - P_jf(y)],1\rangle_{{\cal H}_{N-1}}= 0$ for each $y$ and $j$,
$$\eqalign{\langle [f_{j,y} - P_jf(y)],Q
[f_{j,y} - P_jf(y)]\rangle_{{\cal H}_{N-1}} &\le \lambda_{N-1}\| f_{j,y} - P_jf(y)\|^2_{{\cal H}_{N-1}}\cr
 &= \lambda_{N-1}\left(\|f_{j,y}\|^2_{{\cal H}_{N-1}} - |P_jf(y)|^2_{{\cal H}_{N-1}}\right)\ .\cr}$$
Averaging over $j$ and integrating over $y$,
$${1\over N}\sum_{j=1}^N\int_{Y_N}\
\langle [f_{j,y} - P_jf(y)],Q
[f_{j,y} - P_jf(y)]\rangle_{{\cal H}_{N-1}}
{\rm d}\nu_{N}(y) \le \lambda_{N-1}\left(\|f\|^2_{{\cal H}_N} - \langle f, Pf \rangle_{{\cal H}_N}\right)$$
{}From this and \eqv(mid), \eqv(basrec) follows, since
$f$ itself is an admissible trial function for $\mu_N$.  
The final statement is an evident consequence of the proof of
\eqv(basrec). \eop
\bigskip 
\noindent{\bf Corrolary 2.3:} {\it With $\kappa_N$ and $\beta_N$ defined as in \eqv(kadef) 
and \eqv(bedef), define
$$\Delta_N = N(1 - \lambda_N)\ .\Eq(bdeldef)$$
Then
$$\Delta_N \ge (1 - \max\{\kappa_N,\beta_N\})\Delta_{N-1}\Eq(bdrec)$$
for all $N\ge 3$,
and hence for all $N>2$,
$$\Delta_N \ge \prod_{j=3}^N (1 - \max\{\kappa_j,\beta_j\})\Delta_{2}\ .\Eq(infprod)$$
}
\bigskip
\noindent{\bf Proof:} This follows  directly from \eqv(muform), \eqv(basrec) and \eqv(bdeldef). \eop
\bigskip
We see that a sufficient condition for $\liminf_{N\to \infty}\Delta_N > 0$ is $\Delta_2>0$ and
$$\prod_{N=3}^\infty (1 - \max\{\kappa_N,\beta_N\}) > 0\ .$$
Assuming that $\max\{\kappa_N,\beta_N\} < 1$ for all $N>3$, this last condition is of course
satisfied whenever
$$\sum_{N=3}^\infty \max\{\kappa_N,\beta_N\} < \infty\ .$$

\bigskip

\chap {3: Analysis of the Kac Model }3

\bigskip

The Kac model, with $(X_N,{\cal S}_N,\mu_N)$ being $S^{N-1}$ equipped with its 
rotation invariant probability measure and 
$$Qf(v_1,v_2,\dots,v_N) = 
{\ncht}^{-1}\sum_{i<j}^N\int_{-\pi}^{\pi}\rho(\theta)
f(v_1,v_2,\dots,v^*_i(\theta),\dots,v^*_j(\theta),\dots,v_n){\rm d}\theta\ ,$$
was the basic motivating example for the definition of a Kac system made in the previous
section, where the rest of the elements of the system, namely the action of $\Pi_N$,
the spaces $(Y_N,{\cal T}_N,\nu_N)$, and the maps $\pi_j$ and $\phi_j$ have all been specified.

All that remains to be done before we apply the results of section 2 is to compute the spectrum of
$K$. There are a number of ways that this can be done. The method presented here is the one that most readily adapts 
to the case of three dimensional momentum conserving collisions, which we treat in the next section.
In a later section we shall use a more
group theoretic approach when we discuss the generalization of the Kac walk to $SO(N)$.

\bigskip
\noindent{\bf Theorem 3.1} {\it There is a complete orthonormal set $\{g_n\}$, 
$n\ge 0$,
of eigenfunctions of $K$ where $g_n$ is a polynomial of degree $n$ and the 
corresponding eigenvalue $\alpha_n$ is zero if $n$ is odd, and if $n=2k$,
$\alpha_n$ is given by
$$\alpha_{2k} = (-1)^k{\left|S^{N-3}\right|\over \left|S^{N-2}\right|}
\int_0^\pi(1-\sin^2\theta)^k\sin^{N-3}\theta{\rm d}\theta$$
 In particular, 
 $$\eqalign{
\alpha_2 &= -{1\over N-1}\cr
\alpha_4 &= {3\over N^2-1}\cr
\alpha_6 &= {15\over (N-1)(N+1)(N+3)}\cr 
\alpha_8 &= {105\over  (N-1)(N+1)(N+3)(N+5)}\cr}\Eq(avals)$$
and  $|\alpha_{2k+2}| < |\alpha_{2k}|$ for all $k$. Hence for the Kac model,
$$\max\{\kappa_N,\beta_N\} = \kappa_N = {3\over N^2 -1}\ .$$}
\bigskip

\noindent{\bf Proof:} We have already deduced an explicit form \eqv(kack) for $K$ in the previous section.
We note that by an obvious change of variable, we may rewrite it as
$$Kg(v) = {\left|S^{N-3}\right|\over \left|S^{N-2}\right|}
\int_0^\pi g(\sqrt{1-v^2}\cos\theta)\sin^{N-3}\theta{\rm d}\theta\ .$$
The right hand side is clearly an even function of $v$. Since the operator $K$
preserves parity, it evidently annihilates all odd functions. Hence we may
assume that $g$ is even.

Further, since $\left(\sqrt{1-v^2}\right)^{2k} = (1-v^2)^k$ 
is a polynomial of degree $2k$ in $v$, we see that the space of polynomials of
degree $2n$ or less is invariant under $K$ for all $n$. This implies that
the eigenvectors are even polynomials, and that there is exactly one such 
eigenvector for each degree $2k$.

Now let $g_{2k}$ be the eigenvector that is a polynomial of degree $2k$,
and let $\alpha_{2k}$ be the corresponding eigenvalue. We may normalize
$g_{2k}$ so that the leading coefficient is $1$, and we then have
$$g_{2k} = v^{2k} + h(v)$$
where $h(v)$ is an even polynomial in $v$ of degree no more than $2k-2$.
Thus 
$$\alpha_{2k}v^{2k} + \alpha_{2k}h(v) = 
\alpha_{2k}g_{2k} = Kg_{2k} = Kv^{2k} + Kh(v)$$
This implies that
$$Kv^{2k} = \alpha_{2k}v^{2k} + {\rm lower\ order}\ .$$
The result now follows directly from the formula for $K$, 
the recurrence relation
$$\int_0^{\pi}\sin^n(\theta){\rm d}\theta = {n-1\over n}\int_0^{\pi}\sin^{n-2}(\theta){\rm d}
\theta  \ , \Eq(sinrec) $$
and the fact that $K1 = 1$.
Observe
that the leading coefficient of $v$ in $(1-v^2)^k$ is $(-1)^k$.\quad \eop
\medskip

It is evident from \eqv(avals) that, using the notation of Theorem 2.1, $\kappa_N = 3/(N^2-1)$
and $\beta_N= 1/(N-1)^2$. Hence, for $N \geq 3$, $\kappa_N >\beta_N$, and Theorems 1.1 and 1.2 are now proved.

In order to prove Theorem 1.3, it is necessary to determine $\Delta_2$. 
But in \eqv(AQ2y) we have already determined
$\lambda_2$, and since $\Delta_2 = 2(1-\lambda_2)$, it follows that
$$\Delta_2 =2 \inf_{k\ne 0}
\left\{\int_{-\pi}^{\pi}(1-\cos(k\theta))\rho(\theta){\rm d}\theta\right\}\ .$$
By the Riemann--Lebesgue lemma, $\lambda_2 < 1$, and so in any case $\Delta_2 > 0$. 
In the case Kac considered, $S$ is just the projection onto the constants and
$\lambda_2= 0$ so that $\Delta_2 = 2$.

It remains to solve the recurrence relation \eqv(recA). Notice that
$$1 - \kappa_N = {(N-2)(N+2)\over(N-1)(N+1)}\ .\Eq(goodform)$$
The  product of these terms collapses
and 
$$\prod_{j=3}^N{(j-2)(j+2)\over(j-1)(j+1)} = {1\over 4}{N+2\over N-1}\ .\Eq(finprod) $$
Hence
$$\prod_{j=3}^\infty{(j-2)(j+2)\over(j-1)(j+1)} = {1\over 4}\ ,\Eq(infprod2) $$
and it then follows from \eqv(infprod) of Corollary 2.3 that
$$\Delta_N \ge {1\over 4}{N+2\over N-1}\Delta_2 = 
{1\over 4}{N+2\over N-1}2(1 - \lambda_2) = {1 - \lambda_2\over 2}{N+2\over N-1}\ .\Eq(lowerb)$$

Now, we inquire into the sharpness of this result. By Theorems 2.1 and 3.1,
$$Pf_N = \mu_N f_N\Eq(psat)$$
if and only if $f_N$ has the form $f_N = \sum_{j=1}^Ng_N\circ\pi$ and $Kg_N = (3/(N^2-1))g_N$. That is,
\eqv(psat) holds exactly when, up to a multiple of
$$f_N(\vec v) = \sum_{j=1}^N\left(v_j^4 - \langle 1, v_j^4\rangle\right)\ .$$
By the last part of Theorem 2.2, the bound obtained in Theorem 1.3 can only 
be sharp if $Qf_N = \lambda_N f_N$ for each
$N$. Hence it is natural to compute
$Qf_N$. The result is contained in the next lemma.
\medskip
\noindent{\bf Lemma 3.2} {\it For $f_N(\vec v) = \sum_{j=1}^N\left(v_j^4 - \langle 1, v_j^4\rangle\right)$,
$$Qf_N = \left(1 - {2\gamma (N+2)\over N(N-1)}\right)f_N\Eq(possev)$$
where
$$\gamma = {1\over 4}\left(1 - \int_{-\pi}^{\pi}\cos(4\theta)\rho(\theta){\rm d}\theta\right)\ .\Eq(gamval)$$}
\medskip
\noindent{\bf Proof:} This is a straightforward calculation. \eop
\medskip
Clearly, for the original Kac model, with $\rho$ uniform, $\gamma = 1/4$, and so \eqv(possev) implies that
$\Delta_N = N(1 - \lambda_N)$ is no larger than $(N+2)/(2(N-1))$. Since for the original Kac model $\lambda_2 = 0$,
this upper bound on $\Delta_N$ coincides with the lower bound in \eqv(lowerb) and hence \eqv(lowerb) is sharp in this case.

In fact, the upper bound on $\Delta_N$ provided by Lemma 2.3 coincides with the lower bound in \eqv(lowerb)
whenever
$f_2(v_1,v_2) = v_1^4+v_2^4 - (3/4)$ is  such that
$$Qf_2 = \lambda_2f_2\ .\Eq(twop)$$
Writing $v_1 = \cos(\theta)$ and $v_2 = \sin(\theta)$, we have
$$f( \cos(\theta), \sin(\theta)) = {\cos(4\theta)\over 4}\ .$$
Hence \eqv(twop) certainly holds whenever \eqv(4cond) holds.
Finally, the fact that under the condition \eqv(4cond), $f_N$ is, up to a multiple, the only eigenfuncton of $Q$ with
eigenvalue
$\lambda_N$ follows directly from Theorems 3.1, which says $\kappa_N$ has multiplicity one, and
Theorem 2.1. This completes the proof
of Theorem 1.3.

We shall show in section 7 of this paper that actually in a wide range of circumstances
$$Qf_N = \lambda_Nf_N$$
for all $N$ sufficiently large, even if this is false for, say,  $N=2$. Thus in a great many cases Lemma 3.2
provides the precise value of $\lambda_N$, and hence $\Delta_N$, for large $N$. However, before returning
to analyze the Kac model in this detail, we proceed to give several more examples of Kac systems.

Having explained how our exact determination of the gap for Kac's original model works it is appropriate
to compare this approach with Janvresse's [\rcite{J}] application of Yau's martingale method 
[\rcite{Y1}], [\rcite{Y2}] to the same problem. 
There are 
similarities between our analysis and Yau's method,
in that Yau's martingale method uses induction on $N$, correlation estimates, and the same
conditional expectation operators $P_j$. There are, however, significant differences, as 
indicated
by the difference between Janvresse's estimate and our exact calculation.

First, in Yau's method the spectrum of the $P_j$ operators is estimated not in $\hh$, but in 
the Hilbert space whose inner product is $\langle h,(I-Q)h\rangle$, the so--called Dirichlet 
form space associated to $Q$. This means that the details of the dynamics enter (through Q)
at each stage of the induction, while in our approach purely geometric estimates, 
as described in Theorem 1.1, relate  $\Delta_N$ to $\Delta_{N-1}$.

Second, Yau's method was
designed to handle problems without the permutation symmetry that is present in 
the class of models considered here. The   method  just described
makes full use of this symmetry.
As an example, using this symmetry, we need
only to produce spectral estimates on $P$, the average of the $P_j$.
That the inductive argument presented here
makes full use of this permutation symmetry is one source of its incisiveness in 
this class of problems.

\bigskip

\chap {4: Analysis of the Boltzmann Collision Model }4

\bigskip

Consider now a pair of identical particles with velocities $v_i$ and $v_j$ in 
$\R^3$.
Now we will require that the collisions conserve momentum as well as energy.
These are four constraints on six variables, and hence the set of all kinematically possible 
collisions is two dimensional. It may be identified with $S^2$ as follows:
For any unit vector $\omega$ in $S^2$, define
$$v^*_i(\omega) = v_i + (\omega\cdot(v_j - v_i))\omega\Eq(C1)$$
$$v^*_j(\omega) = v_j - (\omega\cdot(v_j - v_i))\omega\Eq(C2)$$
Now specify $N$ velocities $\vec v = (v_1,v_2,\dots,v_N)$ before the collision with
$$\sum_{j=1}^N |v_j|^2 = E\qquad {\rm and}\qquad \sum_{j=1}^N v_j =  0\ .\Eq(C3)$$
The random collision mechanism is now that we pick a pair ${i,j}$, $i<j$, uniformly at random,
and then pick an $\omega$ in $S^2$ at random, and the post--collisonal 
velocities then become
$$((v_1,\dots,v^*_i(\omega), \dots,v^*_j(\omega),\dots,v_N)\ .$$
We then define the one step transition operator $Q$ by
$$Q_f(\vec v) = 
{\nmcht}^{-1}\sum_{i<j}^N\int_{S^2}
f(v_1,v_2,\dots,v^*_i(\omega),\dots,v^*_j(\omega),\dots,v_n)b(\omega\cdot(v_i-v_j)/|v_j-v_j|){\rm d}\omega\ ,
\Eq(bqdef)$$
where $b$ is a non-negative function on $[-1,1]$ so that 
$$2\pi\int_0^\pi b(\cos\theta)\sin(\theta){\rm d}\theta = 1\ .$$
The function $b$ puts a weight on the choice of $\omega$ so as to determine the relative likelihood of
various scattering angles. 
This definition differs from corresponding definition for the Kac model chiefly through the more
complicated formulae \eqv(C1) and \eqv(C2) paramaterizing three dimensional momentum conserving collisions.
We begin the analysis of this Boltzmann collision model by specifying the structure needed 
to display it as a Kac system.

By choice of scales and coordinates, we may assume that 
$$\sum_{j=1}^N|v_j|^2 = 1\qquad{\rm and}\qquad \sum_{j=1}^Nv_j =0\ \Eq(bcon)$$
both hold initially, and hence for all time. Thus our state space $X_N$ is the set of
all vectors 
$$\vec v = (v_1,v_2,\dots,v_N)\in \R^{3N}$$ satisfying the constraints in \eqv(bcon).
We equip $X_N$ with its Borel field and the metric and uniform probability measure
inherited from its natural embedding in $\R^{3N}$. The symmetric group $\Pi_N$ acts 
on $X_N$ as follows: for $\sigma\in \Pi_N$, 
$$\sigma(v_1,v_2,\dots,v_N) = (v_{\sigma(1)},v_{\sigma(2)},\dots,v_{\sigma(N)})\ .$$
This action is clearly measure preserving. We note that $X_N$ is geometrically equivalent
to the unit sphere $S^{3N-4}$ in $\R^{3N-3}$, but apart from identifying normalization 
factors in our probability measures, this identification is not conducive
to  efficient computation because any embedding in $\R^{3N-3}$ obscures the action of the symmetric group.

To identify the single particle state space $Y_N$, note that
$$\sup\{|v_N|\ |\ (v_1,v_2,\dots,v_N)\in X_N\ \} = {N-1\over N}\ .\Eq(maxrad)$$
To see this, fix $v_N$ and observe that due the momentum constraint in \eqv(bcon),
$\sum_{j=1}^{N-1}v_j= -v_N$. To maximize $|v_N|$, we must minimize the energy in the first $N-1$ particles.
However, by convexity it is clear that
$$\inf\left\{\sum_{j=1}^{N-1}|v_j|^2\ \bigg|\ \sum_{j=1}^{N-1}v_j= -v_N \ \right\}$$
is attained at
$$(v_1,v_2,\dots,v_{N-1}) = -{1\over N-1}(v_N,v_N,\dots, v_N)\ ,$$
which leads directly to \eqv(maxrad).

In short, the momentum constraint prevents all of the energy from belonging to a single particle,
and so each $v_j$ lies in the ball of radius $\sqrt{(N-1)/N}$ in $\R^3$. (While this is true for $N=2$,
this case is somewhat special. For $N=2$, $v_2 = - v_1$ and so $|v_2| 
= 1/\sqrt 2$, rather than 
$|v_2| \le 1/ \sqrt 2$.)

We could take $Y_N$ to be the ball of radius $\sqrt{(N-1)/N}$ in $\R^3$, for $N>3$,
which would then depend on $N$. However,
certain calculations will work out more simply if we rescale and take 
$Y_N$ to be the unit ball in $\R^3$, independent of
$N$. Therefore, we define, for $N \geq 3$,
$$Y_N = \{ v\in \R^3\ |\ |v| \le 1\ \}$$
and let ${\cal T}_N$ be the corresponding Borel field. We take $Y_2$ to be the unit sphere in $\R^3$.
We are then led to define $\pi_j:X_N\rightarrow Y_N$ by
$$\pi_j(v_1,v_2,\dots,v_N) = \left({N\over N-1}\right)^{1/2}v_j\ .\Eq(bpidef)$$
The measure $\nu_N$ is now determined through \eqv(push), but before deducing an explicit formula for it,
we introduce the maps $\phi_j:X_{N-1}\times Y_N\rightarrow X_N$, through which this formula is readily determined.

Consider any fixed $N\ge 3$, so that $X_{N-1}$ is non empty.
Fix a point $\vec w = (w_1,w_2,\dots,w_{N-1})\in X_{N-1}$, and a point $v\in Y_N$. In order that we have
$$\pi_N(\phi_N(\vec w,v)) = v\ ,$$ the $N$th component of $\phi_N(\vec w,v)$
must be $\sqrt{(N-1)/N}v$. Now observe that for any $\alpha\in \R$,
$$\vec v = (v_1,v_2,\dots,v_N) = \left(\alpha w_1 - {1\over \sqrt{N^2-N}}v, \dots,\alpha w_{N-1} -
{1\over \sqrt{N^2-N}}v,  \sqrt{N-1\over N}v\right)$$
satisfies $\sum_{j=1}^Nv_j =0$, and 
$$\sum_{j=1}^N|v_j|^2 = \alpha^2 + |v|^2\ ,$$
since $\sum_{j=1}^{N-1}|w_j|^2 = 1$ and $\sum_{j=1}^{N-1}v_j =0$. Therefore, define
$$\alpha^2(v) = 1 - |v|^2\Eq(alphdef)$$
and
$$\phi_N((w_1,w_2,\dots,w_{N-1}),v) =
\left(\alpha(v) w_1 - {1\over \sqrt{N^2-N}}v, \dots,\alpha(v) w_{N-1} -
{1\over \sqrt{N^2-N}}v,  \sqrt{N-1\over N}v\right)\ ,\Eq(bphindef)$$
and we have that $\phi_N:X_{N-1}\times Y_N\rightarrow X_N$. For $j= 1,\dots,N-1$, let 
$\sigma_{j,N}$ be the pair permutation exchanging $j$ and $N$, and define $\phi_j = \sigma_{j,N}\circ \phi_N$.
We now show that with these definitions \eqv(tens) holds, and in the process, obtain an explicit formula for
$\nu_N$.
\medskip
\noindent{\bf Lemma 4.1} {\it  For $N\ge 3$, the measure $\nu_N$ induced on $Y_N$  through
\eqv(push) for the Boltzmann collision model is
$${\rm d}\nu_N(v) = {|S^{3N-7}|\over|S^{3N-4}|}(1 -|v|^2)^{(3N-8)/2} {\rm d}v\ .\Eq(bnuf)$$
In the case $N=2$, $\nu_2$ is the uniform probability measure on $S^2 = Y_2$.
Moreover, for  these measures $\nu_N$, and with $\phi_j$ defined as above, \eqv(tens)
holds for the Boltzmann collision model for all $N\ge 3$.}
\bigskip
\noindent{\bf Proof:} The measure $\mu_N$ is defined through the natural embedding of $X_N$ in
$\R^{3N}$, and hence it is advantageous  to consider the tangent spaces to 
$X_N$ as subspaces of $\R^{3N}$. Making this
identification, a vector $\vec \xi = (\xi_1,\xi_2,\dots,\xi_N)$ is tangent to $X_N$ at 
$\vec v = (v_1,v_2,\dots,v_N)\in X_N$ provided
$$\sum_{j=1}^N \xi_j\cdot v_j =0\qquad{\rm and}\qquad \sum_{j=1}^N \xi_j = 0\ .\Eq(tan1)$$
Likewise, a vector  $\vec \eta = (\eta_1,\eta_2,\dots,\eta_{N-1})$ is tangent to $X_{N-1}$ at 
$\vec w = (w_1,w_2,\dots,w_{N-1})\in X_{N-1}$ provided
$$\sum_{j=1}^{N-1} \eta_j\cdot v_j =0\qquad{\rm and}\qquad \sum_{j=1}^{N-1} \eta_j = 0\ .\Eq(tan2)$$
And finally, it is clear that the tangent space at any point $v$ of  $Y_N$ is $\R^3$.

Now let 
$$(\phi_N)_*: T_*(X_{N-1})\times T_*(Y_N)\rightarrow T_*(X_N)$$
be the tangent bundle map induced  by $\phi_N$. One easily computes the derivatives and finds that
for a tangent vector $(\vec\eta,0)$ at $(\vec w,v)$,
$$(\phi_N)_*(\vec\eta,0) = (\alpha(v)\vec\eta, 0)\ .\Eq(tan3)$$
Likewise, for a tangent vector $(\vec 0,u)$ at 
 $(\vec w,v)$,
$$(\phi_N)_*(\vec 0,u) = \left({v\cdot u\over \alpha(v)} w_1 - {1\over \sqrt{N^2-N}}u, \dots,
{v\cdot u\over \alpha(v)} w_{N-1} - {1\over \sqrt{N^2-N}}u,  \sqrt{N-1\over N}u\right)\ .\Eq(tan4)$$

Now let $\zeta_X$ be any vector of the type in \eqv(tan3), and let $\zeta_Y$ be any vector of the type
in \eqv(tan4). Obviously
$$\langle \zeta_X,\zeta_Y\rangle  = 0\Eq(ortho)$$
where the inner product is the standard inner product in $\R^{3N}$. Moreover,
$$\langle \zeta_X,\zeta_X\rangle  = 
\alpha^2(v)\langle \vec\eta, \vec\eta\rangle\Eq(xstr)$$
where the inner product on the right is the standard one in $\R^{3N-3}$. 
The determinant of the quadratic form $q_X$ given by
$$\vec\eta \mapsto \alpha^2(v)\langle \vec\eta, \vec\eta\rangle$$
is 
$$\det(q_X) = \alpha^{2(3N-7)}(v)\Eq(det1)$$
since $X_{N-1}$ is $3N-7$ dimensional.
Finally,
$$\langle \zeta_Y,\zeta_Y\rangle  = {(v\cdot u)^2\over \alpha^2(v)} + |u|^2\ ,\Eq(ystr)$$
and the determinant of the quadratic form $q_Y$ given by
$$u \mapsto {(v\cdot u)^2\over \alpha^2(v)} + |u|^2$$
is 
$$\det(q_Y) = 1 + {|v|^2\over \alpha^2(v)} = {1\over \alpha^2(v)}\ .\Eq(det2)$$

Now let $\tilde \mu_N$ and $\tilde \mu_{N-1}$ denote the {\it unnormalized} measures on $X_N$ and
$X_{N-1}$  given by the Riemannian structures induced by their natural Euclidean embeddings.
If $(x_1,\dots,x_{3N-7})$ is any set of coordinates for $X_{N-1}$, and if $(y_1,y_2,y_3)$ are the obvious Euclidean
coordinates for $Y_N$, then these induce, through $\phi_N$, a system of coordinates on $X_N$. 
(Since $X_{N-1}$ is a sphere, 
up to a set of measure zero, one chart of coordinates suffices.)
The volume element  ${\rm d}\tilde \mu_N(x,y)$ in these coordinates can now be expressed the volume element
${\rm d}\tilde \mu_{N-1}(x)$ using \eqv(ortho) \eqv(det1) and \eqv(det2):
$${\rm d}\tilde \mu_N(x,y) = \alpha^{3N-7}(v){\rm d}\tilde \mu_{N-1}(x){1\over \alpha(v)}{\rm d}y\ .$$
Since we know that
$$\int_{X_N}{\rm d}\tilde \mu_N = |S^{3N-4}|\ ,$$
we easily deduce from this that for all continuous functions $f$ on $X_N$,
$$\int_{X_N}f(v){\rm d}\mu_N = 
{|S^{3N-7}|\over|S^{3N-4}|}\int_{Y_N}\left[\int_{X_{N-1}}f\circ \phi_N {\rm d}\mu_{N-1}\right]
(1 -|v|^2)^{(3N-8)/2} {\rm d}v\ .\Eq(finf)$$

Finally, suppose that $f$ has the form $f = g\circ \pi_N$ for some continuous function $g$ on $Y_N$, $N\ge 3$. Then
evidently $f\circ \phi_N(\vec w,v) = g(v)$ everywhere on $X_{N-1}\times Y_N$ and hence
by the definition \eqv(push) and \eqv(finf),
$$\int_{Y_N}g{\rm d}\nu_N = \int_{X_N}f{\rm d}\mu_N = {|S^{3N-7}|\over|S^{3N-4}|}\int_{Y_N}g(v)
(1 -|v|^2)^{(3N-8)/2} {\rm d}v\ .\Eq(flint)$$
Hence we see that \eqv(bnuf) holds, and hence that \eqv(tens) holds for the Boltzmann collision model.
\eop
\medskip
\noindent{\bf Lemma 4.2} {\it\ The Boltzmann collision model, consisting of $(X_N,{\cal S}_N,\mu_N)$,
$(Y_N,{\cal T}_N,\nu_N)$, $\pi_j$, $\phi_j$, $j=1,\dots,N$,
and $Q$ as specified in this section constitute
a Kac system as defined in section 2.}
\medskip
\noindent{\bf Proof:} The properties not already established in Lemma 4.1 are now easily checked using \eqv(bphindef).
\eop
\medskip

Now in order to apply the results of section 2 to this Kac system, we need to determine the spectral properties of
the operator $K$.
The explicit form of  $K$ for the
Boltzmann collision model is easily obtained from \eqv(kker): For all functions $g$ on $Y_N$, the unit ball in in $\R^3$,
and all $N > 3$,
$$\eqalign{
Kg(v) &= \int_{X_{N-1}}g\left(\sqrt{N\over N-1}\left(\sqrt{1-|v|^2}w_{N-1} - {1\over
\sqrt{N^2-N}}v\right)\right) {\rm d}\mu_{N-1}(w)\cr
&= \int_{Y_{N-1}}g\left(\sqrt{N\over N-1}\sqrt{1-|v|^2}\sqrt{N-2\over N-1}y - {1\over N-1}v\right) 
(1 -|y|^2)^{(3N-11)/2} {\rm d}\nu_{N-1}(y)\cr 
&= {|S^{3N-10}|\over|S^{3N-7}|}\int_{|y|\le 1}g\left({\sqrt{N^2-2N}\over N-1}
\sqrt{1-|v|^2}y - {1\over N-1}v\right)  (1 -|y|^2)^{(3N-11)/2} {\rm d}y
\ .\cr}\Eq(bkfor)$$
(The restriction to $N>3$ is because \eqv(flint) only gives us the right form for $\nu_N$ is this range.
Indeed, $3N-11$ is negative for $N=3$. The correct analogs of \eqv(flint) and  \eqv(bkfor) are
easily worked out by the same sort of analysis. We do not do this here, as we 
do not need these formulae.)

Several properties of $K$ are evident from \eqv(bkfor). First, $K$ commutes with rotations in
$\R^3$. That is, if $R:\R^3\rightarrow \R^3$ is a rotation, then clearly
$$K(g\circ R) = (Kg)\circ R\ .$$
Hence we may restrict our search for eigenfunctions $g$ of $K$ to functions of the form
$$g(v) = h(|v|)|v|^\ell {\cal Y}_{\ell,m}(v/|v|)$$
for some function $h$ on $\R_+$, and some spherical harmonic ${\cal Y}_{\ell,m}$.

Second, for each $n\ge 0$, $K$ preserves the space of polynomials of degree $n$. To see this notice that
any monomial in $\sqrt{1-|v|^2}w$ that is of odd degree is annihilated
when integrated against $(1 -|w|^2)^{(3N-11)/2} {\rm d}w$,
and any even monomial in $\sqrt{1-|v|^2}w$ is a polynomial in $v$.

Combining these two observations, we see that $K$ has a complete basis of eigenfunctions of the form
$$g_{n,\ell,m}(v) = h_{n,\ell}(|v|^2)|v|^\ell {\cal Y}_{\ell,m}(v/|v|)\Eq(evcomb)$$
where $h_{n,\ell}$ is a polynomial of degree $n$. 

A third observation leads to an explicit
identification of these polynomials and a formula for the eigenfunctions:
Suppose that $Kg(v) = \lambda g(v)$. Let ${\hat e}$ be any unit vector in $\R^3$. Then since $g$ is a polynomial and hence
continuous,
$$\eqalign{
\lim_{t\to 1}Kg(t{\hat e}) &= 
\lim_{t\to 1}{|S^{3N-10}|\over|S^{3N-7}|}\int_{Y_{N-1}}g\left({\sqrt{N^2-2N}\over N-1}\sqrt{1-t^2}w - 
{1\over N-1}{\hat e}\right) (1 -|w|^2)^{(3N-11)/2} {\rm d}w\cr 
&= g\left(-{1\over N-1}{\hat e}\right)\ ,\cr}$$
since $K1 = 1$. Combining this with $Kg(v) = \lambda g(v)$, we have
$$\lambda g({\hat e}) = g\left(-{1\over N-1}{\hat e}\right)\ .\Eq(evrel)$$

Now consider any eigenfunction $g_{n,\ell,m}$ of the form given in \eqv(evcomb), and let $\lambda_{n,\ell}$
be the corresponding eigenvalue, which will not depend on $m$. Then taking any ${\hat e}$ so that ${\cal Y}_{\ell,m}(\hat
e)\ne 0$, we have from
\eqv(evrel) that
$$\lambda_{n,\ell} = {h_{n,\ell}(1/(N-1)^2)\over h_{n,\ell}(1)}\left(-{1\over N-1}\right)^\ell\ .\Eq(bevrat)$$

Finally, a fourth elementary observation identifies the polynomials $h_{n,\ell}$. For all distinct positive integers
$n$ and $p$, the eigenfunctions $g_{n,\ell,m}$ and $g_{p,\ell,m}$ are orthogonal in ${\cal K}_N$. Hence for
each $\ell$, and for
$n \ne p$,
$$\int_{|v|\le 1} h_{n,\ell}(|v|^2)h_{p,\ell}(|v|^2)(1 - |v|^2)^{(3N-8)/2}|v|^{2\ell}{\rm d}v = 0\ .$$
Taking $r = |v|^2$ as a new variable, we have
$$\int_0^1 h_{n,\ell}(r)h_{p,\ell}(r)(1 - r)^{(3N-8)/2}r^{\ell+1/2}{\rm d}r = 0\ .$$
This is the orthogonality relation for a family of Jacobi polynomials in one standard form, and this identifies
the polynomials $ h_{n,\ell}$. A more common standard form, and one that is used in the sources to which we shall refer,
is obtained by the change of variable $t = 2r-1$, so that the $t$ ranges over the interval $[-1,1]$.
Then for $\alpha,\beta >-1$, $J_n^{(\alpha,\beta)}(t)$ is the  orthogonal $n$th degree polynomial for the weight
$(1-t)^\alpha(1+t)^\beta$.
Then with the variables $t$ and $|v|^2$ related as above; i.e.,
$$t = 2|v|^2 - 1\ ,\Eq(tvch)$$
$$h_{n,\ell}(|v|^2) = J_n^{(\alpha,\beta)}(t)\Eq(jaqrel)$$
for 
$$\alpha = {3N-8\over 2}\qquad{\rm and}\qquad \beta = \ell+ {1\over 2}\ .\Eq(albedef)$$

The particular normalization of the Jacobi polynomials is irrelevant here, as we shall be concerned with ratios
of the form $J_n^{(\alpha,\beta)}(t)/J_n^{(\alpha,\beta)}(1)$. Indeed, notice that from \eqv(tvch)
when $|v|^2=1$, $t=1$, and when $|v|^2 = 1/(N^2-N)$, $t = -1 + 2/(N^2-N)$. Hence from \eqv(jaqrel)
and \eqv(bevrat), we see that
$$\lambda_{n,\ell} = 
{J_n^{(\alpha,\beta)}(-1+2/(N^2-N))\over J_n^{(\alpha,\beta)}(1)}
\left(-{1\over N-1}\right)^{\ell}\ .\Eq(Jratio)$$

We summarize this in the following lemma:

\medskip

\noindent{\bf Lemma 4.3} {\it 
Define the functions 
$$g_{n,\ell,m}(v) = h_{n,\ell}(|v|^2)|v|^\ell {\cal Y}_{\ell,m}(v/|v|)
$$
$n\ge 0$, $\ell\ge 0$ and $-\ell\le m\le \ell$,
where the ${\cal Y}_{\ell,m}$ are an orthonormal 
family of spherical harmonics, and the $h_{n,\ell}$ are polynomials
expressible in terms of the Jacobi polynomials through \eqv(jaqrel).
Then $$\{g_{n,\ell,m}\ | \ n\ge 0, \ell\ge 0, -\ell \le m \le \ell\ \}$$
is a complete orthonormal basis of eigenfuntions of $K$. Moreover, if 
$\lambda_{n,\ell}$ is the corresponding eigenvalue, then \eqv(Jratio) holds.}

\medskip

The problem of determining the spectral gap for $K$ is thus reduced to the problem of
determining the largest number of the form \eqv(Jratio). 
The following integral representation of ratios of 
Jacobi polynomials, due to Koornwinder [\rcite{Ko}] (see also [\rcite{A}], pp. 31 {\it ff.}), 
is useful in this regard.

For all $-1 \le x \le 1$, all $n$ and all $\alpha>\beta$,
$${J_n^{(\alpha,\beta)}(x)\over J_n^{(\alpha,\beta)}(1)} = 
\int_0^{\pi}\int_0^1\left[{1+x-(1-x)r^2\over 2} + i\sqrt{1-x^2}r\cos(\theta)\right]^n
{\rm d}m_{\alpha,\beta}(r,\theta)\Eq(Koor)$$
where
$$m_{\alpha,\beta}(r,\theta) = c_{\alpha,\beta}(1-r^2)^{\alpha-\beta-1}
r^{2\beta+1}\left(\sin\theta\right)^{2\beta}{\rm d}r{\rm d}\theta\ ,$$
and $c_{\alpha,\beta}$ is a normalizing constant that 
makes ${\rm d}m_{\alpha,\beta}$ a probability measure.

Notice from \eqv(albedef) that $\alpha>\beta$ exactly when $2\ell < 3N -9$. Hence we define
$$\ell_0 =  {3N-9\over 2}\ .\Eq(elndef)$$
For all $\ell < \ell_0$, we may use \eqv(Koor) to compute $\lambda_{n,\ell}$.

First, however, observe that
$$\eqalign{
&\left|{1+x-(1-x)r^2\over 2} + i\sqrt{1-x^2}r\cos(\theta)\right|^2 = \cr
&{(1+x)^2\over 4} + {(1-x)^2\over 4}r^2 + {1-x^2\over 2}r^2\cos(2\theta)\le 1\cr
}\Eq(mons)$$
with equality exactly when $r=1$, and $\theta = 0\ {\rm or}\ \pi$.

\medskip

\noindent{\bf Lemma 4.4} {\it For all
$\ell < \ell_0(N)$,
and all $m$,
$$
|\lambda_{n,\ell}| < \mu_{n,\ell} 
$$
where, with $x = -1 + 2/(N-1)^2$,
$$\mu_{n,\ell} = \int_0^{\pi}\int_0^1
\left|{1+x-(1-x)r^2\over 2} + i\sqrt{1-x^2}r\cos(\theta)\right|^{n/2}
{\rm d}m_{\alpha,\beta}(r,\theta)\left(-{1\over N-1}\right)^\ell\ .$$
Moreover, for each $\ell$, $n\mapsto  \mu_{n,\ell}$ is monotone decreasing:
$$\mu_{k,\ell} < \mu_{j,\ell}\qquad{\rm for\ all}\quad k> j\ .$$
}
\medskip
\noindent{\bf Proof:} The montonicity follows directly from \eqv(mons), and the rest is a summary of
the discussion above.
\medskip

We now proceed to calculate the eigenvalues for $\ell+1/2 < (3N-8)/2$
using \eqv(Koor) and \eqv(Jratio). The case $n=0$ is trivial:
$$\lambda_{0, \ell} = \mu_{0, \ell} = \left(-{1\over N-1}\right)^\ell\Eq(enzl)$$
for all $\ell < \ell_0$.
The montonicity in Lemma 4.4 now guarantees that for all $n$ and all $3\le \ell < \ell_0$,
$$|\lambda_{0, \ell}| \le \left({1\over N-1}\right)^3\ .\Eq(most)$$

Next, it is straightforward to calculate $\lambda_{1, \ell}$ and 
$\lambda_{2, \ell}$ using \eqv(sinrec) and the beta integral
$$\int_0^1(1-t)^{a-1}t^{b-1}{\rm d}t = {\Gamma(a)\Gamma(b)\over \Gamma(a+b)}\ .$$
The results for $n=1$ and $n=2$
$$
\lambda_{1, \ell}=\left[ \varepsilon -(1- \varepsilon){2 \ell +3 \over 3N-6}
\right] \left(-{1 \over N-1} \right)^{\ell}\Eq(neq1)
$$
and
$$
\lambda_{2, \ell} = \left[ \varepsilon^2 -{4\ell +10 \over 3N-6}\varepsilon
(1- \varepsilon) + (1- \varepsilon)^2{(2 \ell +5)(2 \ell+3) \over (3n-6)(3N-4)}
\right] \left( -{1 \over N-1} \right)^{\ell} \ ,\Eq(neq2)
$$
where 
$$
\varepsilon = {1 \over (N-1)^2} \ .\Eq(epsAdef)
$$
The eigenvalues
$$
\lambda_{1, \ell} = -\left[ {1\over (N-1)}+ {2 \ell N \over 3(N-1)^2} \right]
\left( -{1 \over N-1} \right)^{\ell}  
$$
are all negative and hence irrelevant for calculating the gap of $K$.
Note that $\lambda_{2,0}$ is asymptotically $5/3N^2$ and otherwise 
$\lambda_{2,\ell}$ is $O(1/N^3)$. 
In fact for all $\ell$ in the specified range with $\ell \geq 3$ 
$$
|\lambda_{n, \ell}| \le \mu_{n,\ell} \le \mu_{0,3} \le  {1 \over N^3} \ .
$$

Finally, a very simple computation provides a constant $C$ independent of $N$ so that
$$\mu_{3,0} \le {C\over N^3}\ ,\qquad  \mu_{2,1} \le {C\over N^3}\ ,\qquad {\rm and}
\quad \mu_{1,2} \le {C\over N^3}\ .\Eq(mubnds)$$
(The cases with $n$ odd are most easily done through estimates on cases with even $n$, For example,
since $\mu_{3,\ell} \le \mu_{4,\ell}^{3/4}$
by H\"older's inequality, it suffices to show that $\mu_{4,\ell} = {\cal O}(1/N^4)$.)
Therefore, again using the montonicity form Lemma 4.4, the only values of $(n,\ell)$ 
with $\ell < \ell_0$ such that
$|\lambda_{n,\ell}|$ is of order $(1/N^2)$ or larger are those for which $n + \ell \le 2$. 
By the computations above, we then
have
$$\sup\{ \ \lambda_{n,\ell}\ |\ n+\ell >0, \quad \ell < \ell _0\ \} = \lambda_{2,0}\Eq(sup4)$$
and
$$\inf\{ \ \lambda_{n,\ell}\ |\  \ell < \ell _0\ \} = \lambda_{1,0}\ .\Eq(inf4)$$

Regarding the restriction $\ell < \ell_0$ in \eqv(sup4) and \eqv(inf4),
it is reassuring to note that for the largest value of $\ell$ 
in this range, the corresponding eigenvalues are 
no larger than $(1/(N-1))^{(3N-11)/2}$. 
This suggests that a fairly crude bound on the part of the spectrum corresponding to $\ell \ge \ell_0$
will suffice to eliminate the restriction on $\ell$ in  \eqv(sup4) and \eqv(inf4).
We shall show that this is the case.

For this purpose we need the integral kernel corresponding to the operator $K$. From \eqv(bkfor) we have
that for all $g\in {\cal K}_N$,
$$\eqalign{&\langle g,Kg\rangle_{{\cal K}_N} =\cr
 &{|S^{3N-10}|\over|S^{3N-4}|}\int_{|y|\le 1}\int_{|v|\le 1}g(v)g(u(y,v))  (1 -|y|^2)_+^{(3N-11)/2}(1 -|v|^2)_+^{(3N-8)/2}
{\rm d}y{\rm d}v\ ,\cr}$$
where
$$u(y,v) = \left({\sqrt{N^2-2N}\over N-1}
\sqrt{1-|v|^2}y - {1\over N-1}v\right)\ .$$
Making the change of variables $(y,v) \rightarrow (u(y,v),v)$ we find
$$(1-|y|^2) ={1\over 1 - |v|^2}\left( 1 - {N-1\over N-2}u^2 - {N-1\over N-2}v^2 - {2\over N-2}u\cdot v\right)$$
which leads to
$$\eqalign{&{|S^{3N-4}|\over|S^{3N-10}|}\left({N^2 -2N\over N^2 -2N + 1}\right)^{3/2}\langle g,Kg\rangle_{{\cal K}_N} =\cr
 &\int_{|y|\le 1}\int_{|v|\le 1}g(v)g(u) 
\left( 1 - {N-1\over N-2}u^2 - {N-1\over N-2}v^2 - {2\over N-2}u\cdot v\right)_+^{(3N-11)/2}
{\rm d}y{\rm d}v\ .\cr}\Eq(symk)$$
We therefore define the kernel $K(u,v)$ by
$$K(u,v) = {|S^{3N-10}|\over|S^{3N-4}|}\left({N^2 -2N+1\over N^2 -2N }\right)^{3/2}
\left( 1 - {N-1\over N-2}u^2 - {N-1\over N-2}v^2 - {2\over N-2}u\cdot v\right)_+^{(3N-11)/2}\ .\Eq(kayker)$$

Now let the $g_{n,\ell,m}$ be the normalized eigenfunctions introduced in Lemma 4.3. Then
$$\eqalign{
\lambda_{n,\ell} &= {1\over 2\ell+1}\sum_{m= -\ell}^\ell \langle g_{n\ell,m}, K g_{n\ell,m}\rangle\cr
&=\int h(|u|)h(|v|)\left[{1\over 2\ell+1}\sum_{m= -\ell}^\ell{\cal Y}_{\ell,m}(u)
{\cal Y}_{\ell,m}(v)\right]K(u,v){\rm d}v{\rm d}w\cr
&=\int h_{n,\ell}(|u|^2)h_{n,\ell}(|v|^2)P_\ell((u\cdot v)/|u||v|)K(v,w){\rm d}v{\rm d}w\cr}\Eq(monz)$$
where, $P_\ell(\cos(\theta))$ is 
the Legendre polynomial of order $\ell$, and it is orthogonal to all other 
polynomials of degree strictly less than $\ell$. 

Now note that the positive part taken in \eqv(kayker) is superfluous
unless the values of $|u|$ and $|v|$ are such that both
$$1 -{N-1\over N-2}v^2 - {N-1\over N-2}w^2 \le {2\over N-2}|v||w| \Eq(C5)$$
and
$$1 -{N-1\over N-2}v^2 - {N-1\over N-2}w^2 \ge -{2\over N-2}|v||w|\ .\Eq(C6)$$
Therfore for all values $|u|$ and $|v|$ that are in the complement of the set
defined by \eqv(C5) and \eqv(C6), the kernel $K(u,v)$
{\it is a polynomial of degree $(3N-11)/2$ in $cos(\theta) = u\cdot v/(|u||v|)$}, provided only that $N$ is
odd so that $(3N-11)/2$ is an integer.
Now if $\ell \ge \ell_0$, then then $\ell > (3N-11)/2$
Hence, for such fixed values of 
$|v|$ and $|w|$, the integral over $\theta$  vanishes, and we may as well 
redefine $K(v,w)$ so that it vanishes on the complement of the set
defined by \eqv(C5) and \eqv(C6).

Hence, going back to \eqv(monz), we have the whenever $N$ is odd and $\ell \ge \ell_0$,
$$\lambda_{n,\ell} = {1\over 2\ell+1}\sum_{m= -\ell}^\ell 
\int_{|u| \le 1} \int_{ |v| \le 1}
\tilde g_{n\ell,m}(u)\tilde g_{n\ell,m}(v)K_A(u,v) du dv \Eq(monz1)$$
where
$$K_A(u,v) = \left({N^2-2N+1\over N^2-2N}\right)^{3/2}{|S^{3N-10}|\over |S^{3N-7}|}
{\left( 1 - {N-1\over N-2}|u|^2 - {N-1\over N-2}|v|^2 - {2\over N-2}u\cdot
v\right)_+^{(3N-11)/2}\over
(1-|v|^2)^{(3N-8)/ 4}(1-|u|^2)^{(3N-8)/4}}1_A(u,v)$$
and where $A$ is the set of points $(u,v)\in \R^6$ satisfying \eqv(C5) and \eqv(C6),
and finally where
$$\tilde g_{n\ell,m}(u) = \left({|S^{3N-7}|\over |S^{3N-4}|}\right)^{1/2}(1-|u|^2)^{(3N-8)/2}g_{n\ell,m}(u)\ .$$
The last definition is such that
$$\int_{|u|\le 1}|\tilde g_{n\ell,m}(u) |^2{\rm d}u = \|g_{n\ell,m}\|_{{\cal K}_N}^2 = 1\ ,$$
and so by \eqv(monz1) and the Schwarz inequality, whenever $N$ is odd and $\ell \ge \ell_0$,
$$\lambda^2_{n,\ell} \le \int_{|u|\le 1}\int_{|v|\le 1}|K_A(u,v)|^2{\rm d}u{\rm d}v\ .\Eq(monz3)$$
This leads directly to the following lemma:
\medskip
\noindent{\bf Lemma 4.5} {\it There is a finite integer $N_0$ such that for all odd integers $N\ge N_0$,
and all $\ell \ge \ell_0$,
$$|\lambda_{n,\ell}| \le \left({1\over N}\right)^{N/2}\ .$$}

\medskip
\noindent{\bf Proof:} On account of \eqv(monz3) and the definition of $K_A$, our task is to estimate
$$\int_A 
{\left( 1 - {N-1\over N-2}|u|^2 - {N-1\over N-2}|v|^2 - {2\over N-2}u\cdot
v\right)_+^{3N-11}\over
(1-|v|^2)^{(3N-8)/ 2}(1-|u|^2)^{(3N-8)/2}}   {\rm d}u{\rm d}v\ .$$
Define the quadratic forms
$$q_\pm (x,y) = {N-1\over N-2}x^2 - {N-1\over N-2}y^2 \pm {2\over N-2}xy$$
on $\R^2$. Notice that
$$\left( 1 - {N-1\over N-2}|u|^2 - {N-1\over N-2}|v|^2 - {2\over N-2}u\cdot
v\right) \le 1 - q_-(|u|,|v|)$$
and that 
$$(1-|v|^2)(1-|u|^2) \ge (1-(|u|^2+|v|^2))\ .$$
Hence the integrand above is no larger than
$$ \left(1 - q_-(|u|,|v|)\right)^{3N-11}\over (1-(|u|^2+|v|^2))^{(3N-8)/ 2} \ .\Eq(tempr1)$$

Next note that the eigenvalues of $q_\pm$ are, in both cases,  $1$ and $N/(N-2)$. Hence
$$(|u|^2+|v|^2) \le q_\pm (|u|,|v|) \le {N\over N-2}(|u|^2+|v|^2)\ .\Eq(qqq)$$ 
Thus, by the first of these inequalities, the ratio in \eqv(tempr1) is no greater than
$$ \left(1 - q_-(|u|,|v|)\right)^{(3N-14)/2}\ .\Eq(tempr2)$$
The conditions \eqv(C5) and \eqv(C6)
can be expressed as
$$q_+(|u|,|v|) \ge 1\qquad{\rm and}\qquad q_-(|u|,|v|) \le 1\ .\Eq(qqa)$$
Combining the first inequality in \eqv(qqa) with the second inequality  in \eqv(qqq) we have
that on $A$, $|u|^2+|v|^2> (N-2)/N$, and hence $q_-(|u|,|v|)> (N-2)/N$ on $A$.
Hence on $A$, the quantity in \eqv(tempr2) is no greater than $(2/N)^{(3N-14)/2}$. The result now easily follows.
\eop
\medskip
Note that on account of this result, increasing $N_0$ if need be, the condition $\ell <\ell_0$ in \eqv(sup4)
and \eqv(inf4) may be dropped, and the estimates remain valid, for all odd integers $N$ with $N\ge N_0$.
Our next task concerning the determination of the spectral properties of $K$ is to remove the condition that $N$ be odd.
\medskip
\noindent{\bf Lemma 4.6} {\it Let $\kappa_N$ and $\beta_N$ be defined for the Boltzmann collision model
as in Theorem 2.1. Then for all $N>3$, 
$$|\beta_N| \le |\beta_{N-1}|\Eq(monzB)$$
and for all $N$ such that $ \kappa_{N-1} < 1/2$,
$$\kappa_N \le {\kappa_{N-1}\over 1 - \kappa_{N-1}}\ .\Eq(monzA)$$
}
\medskip
\noindent{\bf Proof:} First let $g$ satisfy $\|g\|_{{\cal K}_N}=1$ and $Kg = \kappa_Ng$. Then
$$\eqalign{\kappa_N &= \langle g, Kg\rangle \cr
&= \int_{X_N}g\circ \pi_1 g\circ \pi_2 {\rm d}\mu_N\cr
&=\int_{Y_N}\left[\int_{X_{N-1}}(g\circ \pi_1\circ \phi_N)(g\circ \pi_2\circ \phi_N){\rm d}\mu_{N-1}\right]
{\rm d}\nu_N\ .\cr}\Eq(monz4)$$
Now,
$$g\circ \pi_1\circ \phi_N(\vec w,v) = h_v\circ \pi_1(\vec w)$$
where
$$h_v(y) = g\left({\sqrt{N^2-2N}\over N-1}\sqrt{1 -|v|}y - {1\over N-1}v\right)\ .$$
Finally, let $\tilde h_v$ be given by
$$\eqalign{\tilde h_v\circ \pi_1 &=  h_v\circ \pi_1 - \int_{X_{N-1}} h_v\circ \pi_1 {\rm d}\mu_{N-1}\cr
&= h_v\circ \pi_1 - P_N(g\circ \pi_1)\cr
&= h_v\circ \pi_1 - Kg\circ \pi_N\ .\cr}$$
Going back to \eqv(monz4) and using the variational definition of $\kappa_{N-1}$ we have, much as in the proof
of Theorem 2.2,
$$\eqalign{
\kappa_N &\le \kappa_{N-1}\int_{Y_N}\left[\int_{X_{N-1}}|h_v\circ \pi_1|^2{\rm d}\mu_{N-1}\right] {\rm d}\nu_N(v) 
+ (1 -  \kappa_{N-1})\|Kg\|^2_{{\cal K}_N}\cr
&= \kappa_{N-1} + (1 -  \kappa_{N-1})\kappa_N^2\ .\cr}\Eq(monz8)$$
Since $\kappa_{N-1}<1$, this last inequality may be written as $P_2( \kappa_N) \ge 0$ where
$$P_2(x) = x^2 - {1\over 1 -  \kappa_{N-1}}x + {\kappa_{N-1}\over 1 -  \kappa_{N-1}} \ .$$
The polynomial $P_2(x)$ has the roots $x=1$ and $x= {\kappa_{N-1}/(1 -  \kappa_{N-1})} < 1$ and
is negative between these two numbers. Since $\kappa_{N} \le 1$  \eqv(monzA) follows.

The proof of \eqv(monzB) is similar but simpler. Suppose
$g$ satisfy $\|g\|_{{\cal K}_N}=1$ and $Kg = \tilde\beta_Ng$ where
$$\tilde \beta_N = \inf\{ \langle h,Kh\rangle \ |\ \|h\|_{{\cal K}_N}= 1\ \}\ .$$
The analysis that lead to \eqv(monz8) now yields
$$\tilde \beta_N \ge \tilde \beta_{N-1} + (1- \tilde \beta_{N-1})\|Kg\|^2_{{\cal K}_N}$$
which certainly implies \eqv(monzB). \eop
\medskip
We are finally ready to prove the analog of the original Kac conjecture for the Boltzmann collision model:
\medskip
\noindent{\bf Theorem 4.7} {\it For the Boltzmann collision model
$$\liminf_{N\to 0}\Delta_N > 0\ .$$}
\medskip
\noindent{\bf Proof:} We choose $N_0$ large enough so that for all odd integers $N>N_0$, 
$$\kappa_N = \lambda_{2,0} \ ,$$
where $\lambda_{2,0}$ is specified in \eqv(neq2) and \eqv(epsAdef). We can do this
since $\lambda_{2,0} \sim 5/N^2$, and Lemma 4.6 tells us, increasing $N_0$ if need be, that 
$$\kappa_N \le {2\over N^2}$$
for all $N >N_0$. Now by Corollary 2.3,
$$\liminf_{N\to 0}\Delta_N > \prod_{j = N_0+1}^\infty (1 - 2/j^2)\Delta_{N_0}\ .$$
The infinite product is clearly strictly positive, and so it remains to verify that $\Delta_N>0$
for all $N$, and in particular $N = N_0$. 

This may as well be done by a compactness argument since we are not being specific about $N_0$. For 
$1 \le i<j \le N$, define
$$R_{i,j}f(\vec v) = \int_{S^2}
f(v_1,v_2,\dots,v^*_i(\omega),\dots,v^*_j(\omega),\dots,v_n)
b(\omega\cdot(v_i-v_j)/|v_j-v_j|){\rm d}\omega\ ,\Eq(rbqdef)$$
so that by \eqv(bqdef),
$$Q_f(\vec v) = 
{\ncht}^{-1}\sum_{i<j}^NR_{i,j}f(\vec v)\ .\Eq(monz9)$$
This operator is not compact. 
In the case where  $\omega$ is selected uniformly, one easily sees that for any unit vector $\hat e\in \R^3$,
and any odd integer $k$, $f_k(\vec v) = \left(\pi_1(\vec v)\cdot \hat e\right)^k$ is an eigenfunction of $Q$,
with a non-zero eigenvalue independent of $k$.  
In the case of the Kac model this was explicitly observed by Diaconis and Saloff--Coste,
and this may have been clear to Kac when he remarked on the difficulty of showing that $\Delta_N>0$ for the original
Kac model. 

However, consider $Q^{2N}$. Observe from \eqv(rbqdef) and \eqv(monz9) that $Q^{2N}$ is an average over
monomials of degree $2N$ in the operators $R_{i,j}$. Each such monomial enters with the same positive weight,
and each is a contraction on ${\cal H}_N$, since clearly each  $R_{i,j}$
is a contraction on ${\cal H}_N$.

Now one such monomial is
$$A = \left(R_{1,2}R_{2,3}R_{3,4}\dots R_{N,1}\right)\left( R_{N,1}\dots R_{4,3}R_{3,2}R_{2,1}\right)$$
which is positive.
It follows that there is a positive number $a$ so that
$$Q^{2N} = aA + (1 -a)B$$
where $B$ is a self--adjoint contraction on ${\cal H}_N$. ($B$ is the average over the remaining
monomials.) Now it is easy to see that $A$ is compact. 
Since it entails averages over each of the variables,
it has a continuous kernel, and hence is Hilbert--Schmidt. Now 
$$\eqalign{
&\sup\{ \langle f,Q^{2N} f\rangle_{{\cal H}_N} \ | \|f\|_{{\cal H}_N} =1, 
\langle 1,f\rangle_{{\cal H}_N} = 1\ \}\cr
&= 
\sup\{ \langle f,aA + (1-a)B f\rangle \ | \|f\|_{{\cal H}_N} =1, \langle 1,f\rangle_{{\cal H}_N} = 0\ \}\cr
&\le a\sup\{ \langle f,A f\rangle \ | \|f\|_{{\cal H}_N} =1, \langle 1,f\rangle_{{\cal H}_N} = 0\ \} + (1-a)\ .\cr}$$
Now since $A$ is compact, 
$\sup\{ \langle f,A f\rangle_{{\cal H}_N} \ | \|f\|_{{\cal H}_N} =1, \langle 1,f\rangle_{{\cal H}_N}
= 0\ \}$ is attained. And clearly if $f$ satisfies $\|f\|_{{\cal H}_N} =1$, $ \langle 1,f\rangle_{{\cal H}_N} =0$,
and $\langle f,A f\rangle_{{\cal H}_N}=1 $,
then $\|R_{k,k+1}f\|_{{\cal H}_N} = 1 = \|f\|_{{\cal H}_N}$ 
and this is impossible by our ergodicity assumptions. \eop

\bigskip

\chap {5: Analysis of a  Shuffling Model }5

\bigskip

When momentum and energy are conserved for one dimensional velocities,
the only possibility is an exchange of velocities. Thus the Kac walk in this case
is simply a walk on the permutations of $(v_1,v_2,\dots ,v_N)$, which, at least when all of these velocities
are distinct, we may identify with a random walk on the permutation group $\Pi_N$. The corresponding walk has
been throoughly analyzed by Diaconis and Shahshahani [\rcite{DS}], but we briefly discuss it 
in this section to illustrate several
features of our approach. (In fact, they estimate approach to uniformity in the total variation norm, for which they need,
and derive, not only the spectral gap, but information on all of the eigenvalues and their multiplicities.)

Let $X_N = \Pi_N$,  and let 
$$Y_N = \{1,2,\dots, N\}\ .\Eq(E1)$$
For $\sigma\in \Pi_N$, define for $j = 1,2\dots, N$,
$$\pi_j(\sigma) = \sigma(j)\ .\Eq(E2)$$
Let $\mu_N$  be normalized counting measure on $X_N$, so that  $\nu_N$
is normalized counting measure on $Y_N$.

To define the transition function, fix a number $p$ with $0<p<1$, which will represent the probability of
``success''in a coin toss. Consider a deck of $N$ distinct cards which are to be ``shuffled'' as follows:
Pick a pair $i<j$ uniformly at random, and then toss
a coin to generate independent Bernoulli variables with success probability $p$. If the result
of the coin toss is success, exchange cards at the $i$th  and $j$th positions from the top of the deck,
and otherwise do nothing. This procedure is then repeated. 

We can identify the state of the deck at each stage with the
permutation $\sigma$ which puts it in that order starting from a canonical ``unshuffled'' order.
In these terms,  the
current state
$\sigma$ is updated by
$$\sigma \mapsto \sigma_{i,j}\sigma$$
where $\sigma_{i,j}$ is the pair permutation exchanging $i$ and $j$, and fixing all else. If the result is
not success, the current state $\sigma$ is not altered. The one step transition operator
is clearly
$$Qf(\sigma) = \ncht^{-1}\sum_{i<j}\left[pf(\sigma_{i,j}\sigma) + (1-p)f(\sigma)\right]\ .\Eq(E3)$$

To display this as a Kac system, define
$\phi_N: X_{N-1}\times Y_N \rightarrow X_N$ by
$$\phi_N(\sigma, k) = \sigma_{k,N}\tilde \sigma \Eq(E5)$$
where
$\tilde \sigma(j) = \sigma(j)$ for $j \le N-1$, and $\tilde \sigma(N) = N$.
Note that $\pi_N\circ \phi_N(\sigma, k) = \sigma_{k,N}(\tilde \sigma(N)) = 
\sigma_{k,N}(N) = k$. We then define
$\phi_j = \phi_N\circ\sigma_{j,N}$. 
It is clear that 
these maps are bijections, and since $\mu_{N-1}\otimes\nu_N$ is uniform counting measure on 
$X_{N-1}\times Y_N$. \eqv(tens) is trivially true. Thus it is clear that this shuffling model is
a Kac system.

Moreover it is easy to see that
$$Kg(i) = \sum_{j=1}^NK_{i,j}g(j)\Eq(E23)$$
where 
$$K_{i,j} = {1\over N-1}(1 - \delta_{i,j})\ .\Eq(E24)$$
Hence $K$ has the eigenvalues $1$, with multiplicity one, and $-1/(N-1)$ with 
multiplicity $N-1$. Hence for this model, with $\kappa_N$ and $\beta_N$ as in Theorem 2.1,
$$-\kappa_N = (N-1) \beta_N = {1\over N-1}\Eq(E25)$$
and thus by Corollary 2.3
$$\Delta_N \ge \prod_{j=3}^N\left(1 - {1\over (j-1)^2}\right)\Delta_2\ .\Eq(E26)$$
Again, this product collapses, and one finds
$$\prod_{j=3}^N\left(1 - {1\over (j-1)^2}\right) = \left({N\over N-1}\right){1\over 2}\ .\Eq(E27)$$

Clearly $Q_2$ may be identified with the matrix
$$\left[\matrix{ &1-p &p\cr &p &1-p\cr}\right]\Eq(E29)$$
and hence $Q_2$ has the eigenvalues $1$ and $1-2p$. Hence $\lambda_2 = 1 -2p$,
and $\Delta_2 = 4p$. Combining this with
\eqv(E26) and \eqv(E27), we have
$$\Delta_N \ge {N\over N-1}2p\ .\Eq(ro7)$$

To see that this result is sharp, we need to display an appropriate eigenfunction. We know from Theorem 2.2 that 
\eqv(ro7) can be
sharp if and only if there is a function $f_N$  satisfying both $Qf_N = \lambda_Nf_N$ and 
$Pf_N = \mu_N f_N$. Theorem 2.1 then tell us that since $\beta_N > \kappa_N$ for this problem, we get an eigenfunction of
$P$ with $Pf_N = \mu_Nf_N$ from eigenfunctions $h$ of $K$ with $Kh = -1/(N-1)h$ through
$$f_N = h\circ \pi_i - h\circ \pi_j\qquad{\rm for}\quad 1 \le i < j \le N\Eq(ro9)$$
for some $i<j$. A tedious but straightforward computation, using $\sum_{j=1}^N h(j) = 0$, which is equivalent to
$Kh = -1/(N-1)h$, shows that
$$Qf_N =\left(1- {2p\over N-1} \right)f_N\ .$$
This implies that  $\lambda_N \ge 1 - 2p/(N-1)$ and hence
$$\Delta_N \le {N\over N-1}2p\ .\Eq(ro8)$$
This leads to the following result:
\bigskip
\noindent{\bf Theorem 5.1} \ {\it The binary shuffling model is a Kac system, and 
$$\Delta_N = {N\over N-1}2p\ .\Eq(ro17)$$
Moreover,
$$N(I-Q)f = \left({N\over N-1}2p\right)f$$
if and only if $f$ has the form specified in \eqv(ro9) for some function $h$ on $\{1,2,\dots, N\}$
such that $\sum_{j=1}^N h(j) = 0$. In particular, $\lambda_N$ is an eigenvalue of $Q$ of multiplicity $(N-1)^2$.}
\bigskip
\noindent{\bf Proof:} The equality \eqv(ro17) follows from \eqv(ro7) and \eqv(ro8), and this identifies $\lambda_N$. 
We have shown above that 
every function $f$ of the form \eqv(ro9) with $\sum_{j=1}^N h(j) = 0$ satisfies $Qf = \lambda_Nf$, and by Theorem 2.2,
the converse holds as well since any such $f$ must also satisfy $Pf = \mu_Nf$, and this occurs only when $f$
has the specified form. Finally, it is easily seen that the $N-1$ functions
$$h\circ \pi_1 - h\circ \pi_2\ ,\  h\circ \pi_2 - h\circ \pi_3\ ,\ h\circ \pi_3 - h\circ \pi_4\ ,\
\dots \ ,h\circ \pi_{N-1} - h\circ \pi_N\ $$
are a basis for the span of the functions of the form specified in \eqv(ro9) whenever $\|h\|_{\cal K} \ne 0$
and $\sum_{j=1}^N h(j) = 0$. Also if $h$ and $\tilde h$ are any two orthogonal eigenfunctions of $K$,
$h\circ \pi_i - h\circ \pi_j$ is orthogonal to $\tilde h\circ \pi_k - \tilde h\circ \pi_\ell$ for all
$i<j$ and $k<\ell$. Since the $-1/(N-1)$ has multiplicity $N-1$ as an eigenvalue of $K$, the final statement is now
shown. \eop
\bigskip

Diaconis and Shahshahani actually devote most of their attention to
the model in which the success probability $p$ depends on $N$ through $p  =1-1/N$. The present methods are easily
adapted to handle the case in which $p$ depends on $N$. Let $Q_r$ denote the transition operator defined in \eqv(E3).
Then clearly  for two different success probabilities $p$ and $p'$,
$$Q_p = {p\over p'} Q_{p'} + \left({p\over p'} -1 \right)I\ .$$
This may be used to take into account the effects of the $N$ dependence in $p$ on $\lambda_N$.

\bigskip

\chap {6: The Kac Walk on $SO(N)$ }6

\bigskip

Let $SO(N)$ denote the group of orthogonal $N\times N$ matrices with unit determinant. 
In this section we consider a
generalization of the original Kac model in which the state space is $SO(N)$ instead of $S^{N-1}$.
This generalization was
introduced by Diaconis and Saloff--Coste [\rcite{DSC}], and studied by Maslin as well,
both  in the case of ``uniformly selected rotations'', i.e., $\rho(\theta) = 1/2\pi$.
To explain the nature of the underlying process, 
which these authors  call the ``Kac walk on $SO(N)$'',
we  let
$R_{i,j}(\theta)$ denote the same rotation in $\R^N$ that was used in \eqv(A6a), except now we identify it
with the corresponding $N\times N$ matrix, and we will now consider our $N$--dimensional velocity vectors $\vec v$ as
column vectors of an othogonal matrix. Then multiplying 
 $R_{i,j}(\theta)$ and the ``pre--collisional velocity
vector'' $\vec v$ produces the 
``postcollisional velocity vector'', just as in the 
original Kac model.

Given a continuous function $f$ on $SO(N)$, define
$$Qf(G) =  \left(\matrix{N\cr 2\cr}\right)^{-1}\sum_{i<j}\int_{-\pi}^\pi 
f(R_{i,j}(\theta)G)\rho(\theta){\rm d}\theta \Eq(SG6)$$
where $\rho(\theta)$ satisfies the same conditions imposed on $\rho(\theta)$
in the original Kac model. 

The connection with the Kac walk on $S^{N-1}$ 
becomes quite clear when one writes $G$ in terms of its
$N$ columns, $G = [\vec v_1,\vec v_2,\dots,\vec v_N]$, since then 
$$R_{i,j}(\theta)G = [R_{i,j}(\theta)\vec v_1,R_{i,j}(\theta)\vec v_2,
\dots,R_{i,j}(\theta)\vec v_N]\ .$$
Each of the $\vec v_j$ is an element of $S^{N-1}$, and it is clear from \eqv(A6a)
that under the Kac walk on $SO(N)$, each column
of $G$ is a Markov process in its own right, and is in 
fact a copy of the original Kac walk on $S^{N-1}$.
Therefore, if $f$ depends on $G$ only through the first column 
of $G$, which is an element of $S^{N-1}$, $Qf$
coincides with what we would get by applying the $Q$ operator 
for the Kac model to $f$ considered as a function on
$S^{N-1}$. In this sense the Kac walk on $SO(N)$ is a 
generalization, and indeed, an extension, 
of the Kac
walk on
$S^{N-1}$. 

This relation between the Kac walks on $SO(N)$ and $S^{N-1}$ provides an immediate upper bound
on the spectral gap for the walk on $SO(N)$:
We see, by restricting
the class of test functions to those that depend on $G$ only through a single column, that
 the spectral gap for the
Kac walk on $SO(N)$ cannot be larger than the spectral gap for the
Kac walk on $S^{N-1}$. 

In fact, as found by Maslin in the case in which $\rho(\theta)$ is
uniform, the two gaps
actually coincide.
In this section, we prove this also when  $\rho$ is not assumed to be uniform.
The Kac walk on $SO(N)$  provides a good illustration of the methods of this 
paper in which the ``single particle space''
depends on $N$. This example goes beyond our previous examples in 
other ways as well, as we shall see as soon as we begin with displaying
it as a Kac system.

It turns out that it is most convenient to do this through consideration of the Kac
walk on $O(N)$, the group of orthogonal $N\times N$ matrices. For a continuous function $f$ on 
$O(N)$, we define $Qf$ exactly as above, except that now $G$ now ranges over  $O(N)$.
We equip $O(N)$ with
its Borel field and its  normalized Haar
measure $\mu_N$. Then by our assumptions on $\rho$, $Q$ extends to be a self adjoint 
Markovian contraction on
${\cal H}_N = L^2(O(N),\mu_N)$. However, it is not ergodic; the nullspace of $Q$ is
two dimensional, and spanned by $1$ and ${\rm det}$. Of course, on the subspace
$L^2(SO(N),\mu_N)$,  it is ergodic.

The reason for working in the non-ergodic $L^2(O(N),\mu_N)$ setting is that permutation 
symmetry plays an important role in our analysis. The natural action of $\Pi_N$ on $O(N)$
is through interchange of rows. Note that this extends the action of $\Pi_N$ on $S^{N-1}$,
considered as the first column of $G$, that we used in our analysis of the original Kac model.
Interchange of two rows of an element of $SO(N)$ of course changes the sign of the determinant,
and so does not preserve $SO(N)$.
An alternative is to conjugate elements of $SO(N)$ by the permutations; that is, 
to swap both rows and columns. This however complicates the construction of a Kac system
for the model, and in any case, it is of some interest to show that the methods used here
can be applied when there is more than one ergodic component. This said, we proceed with the
construction of the Kac system.

The ``$N$--particle space'' $(X_N,{\cal S}_N,\mu_N)$ will of 
course be $O(N)$ equipped with
its Borel field and its  normalized Haar
measure is $\mu_N$, as indicated above. Let $O(N)_+$ denote the component of
$O(N)$ on which the determinant is positive, so that $O(N)_+$ is just
$SO(N)$, and let 
 Let $O(N)_-$ denote the component of
$O(N)$ on which the determinant is negative.

For any permutation $\sigma$ in $\Pi_N$, let $P_\sigma$ denote 
the corresponding
$N\times N$ permutation matrix. For $G$ in  $O(N)$, define $\sigma(G)$ by
$$\sigma(G) = P_\sigma G\ .$$
That is, $\sigma$ acts on the matrix $G$ by permuting its rows. Clearly this is a measure
preserving action of $\Pi_N$ on $O(N)$.

We take the {\it single particle space} $(Y_N,{\cal T}_N,\nu_N)$ to be $S^{N-1}$ equipped
with its Borel field and
normalized rotation invariant measure $\nu_N$.
For each $j =1,2,\dots,N$, let ${\bf e}_j$ be the $j$th standard basis vector
in $\R^N$, written as a row vector, so that for any $N\times N$ matrix $A$, ${\bf e}_jA$ is
 the $j$th row of $A$.
We then define
$$\pi_j(G) = {\bf e}_jG\ .$$
That is, $\pi_j(G)$ is the $j$th row of $G$.
It is clear that $\pi_j:X_N\rightarrow Y_N$ and that 
$$\pi_j(\sigma(G)) = 
\pi_{\sigma(j)}(G)$$ for each $j$ and $G$.
So far, we have verified the first two features required of a Kac system.

The next steps in the construction of this Kac system are slightly more involved. 
We have to construct the
maps $\phi_j: X_{N-1}\times Y_N \rightarrow X_N$, but since there is no canonical embedding
of $O(N-1)$ into $O(N)$, they have to be constructed ``by hand'', using a convenient coordinate
chart. (Just as with the original Kac model, the maps $\phi_j$ cannot be continuous
since $ X_{N-1}\times Y_N$ and $X_N$ just do not have the same topology. But
just as in that case, we are only concerned with measure theoretic
properties of these mappings, and on a set of full measure they will be well behaved.)

Let
$$\vec v_- = (0,0,\dots,0,-1)$$
denote the ``south pole'' in $S^{N-1}$. We may use the stereographic 
projection to identify $S^{N-1}\backslash \{\vec v_-\}$
with $\R^{N-1}$. At each point of $\R^{N-1}$ we of course have the 
standard orthonormal basis. The stereographic projection,
which is conformal, caries this back to an orthogonal basis for the tangent 
space at the corresponding point in
$S^{N-1}\backslash \{\vec v_-\}$. Normalizing these vectors, we thus obtain a smoothly 
varying frame of orthonormal vectors
$$\{{\bf u}_1(\vec v), {\bf u}_2(\vec v), \dots {\bf u}_{N-1}(\vec v)\}$$
in $\R^N$ parameterized by $\vec v$ in $S^{N-1}\backslash \{\vec v_-\}$. 
For each $\vec v$, they form an
orthonormal basis for the tangent space to $S^{N-1}$ at $\vec v$.

Now for each $\vec v$ in $S^{N-1}\backslash \{\vec v_-\}$, define $U(\vec v)$ to be the $(N-1)\times N$ 
matrix whose $j$th row
is ${\bf u}_j(\vec v)$. Let the $j$th row of $U(\vec v_-)$ be ${\bf e}_j$. The map 
$v\mapsto U(\vec v)$ is now defined on all
of $S^{N-1}$, though of course it is discontinuous at the ``south pole'', $\vec v_-$. 
This, however, will not be a problem.

Now define the map
$$\phi_N:O(N-1)\times S^{N-1}\rightarrow O(N)$$
as follows:
Given $\tilde G$ in $O(N-1)$ and $\vec v$ in $S^{N-1}$, first form the $(N-1)\times N$ matrix $\tilde G U(\vec v)$
 with $U(\vec v)$
as specified above. Because $\tilde G$ is in $O(N-1)$ and the rows of $U(\vec v)$ are orthonormal, 
the rows of
$\tilde G U(\vec v)$ are orthonormal. Moreover, since the rows of $U(\vec v)$ are a basis for the 
tangent space to $S^{N-1}$
at $\vec v$, each one is orthogonal to $\vec v$. The rows of $\tilde G U(\vec v)$ are linear combinations of the
rows of $U(\vec v)$,
and hence these too are orthogonal to $\vec v$. Therefore, if we form the $N\times N$ matrix
$$\left[\matrix{\tilde G U(\vec v)\cr \vec v\cr}\right]$$
by adjoining $\vec v$ to $\tilde G U(\vec v)$ as the final row, we obtain an orthogonal matrix. 

Next observe that the determinant of this matrix is just the determinant of $\tilde G $. 
Indeed, it is clear
that when $\vec v$ is the
``north pole'', so that $U(\vec v)$ consists of the first $N-1$ rows of the 
$N\times N$ identity matrix, then the determinant of this
matrix is simply the determinant of
$\tilde G $, whcih is either  $+1$ or $-1$. Now as $\vec v$ varies in 
$S^{N-1}\backslash \{\vec v_-\}$,
$${\rm det}\left(\left[\matrix{\tilde G U(\vec v)\cr \vec v\cr}\right]   \right)$$
varies continuously. Hence the value is just ${\rm det}(\tilde G )$ for all such $\vec v$.
Continuity fails at the ``south pole'',
but there it is again obvious by the special form of the definition of $\phi_N$ at $\vec v_-$
that still in this case the determinant is still just that of $\tilde G $. Hence the image of
$$(\tilde G ,\vec v) \mapsto \left[\matrix{\tilde G U(\vec v)\cr \vec v\cr}\right] = \phi_N(\tilde G ,\vec v)$$
does indeed lie in $O(N)$, and ${\rm det}(\phi_N(\tilde G ,\vec v)) = {\rm det}(\tilde G )$.
It is also clear by construction that
$$\pi_N(\phi_N(\tilde G ,\vec v)) = \vec v$$
everywhere on $O(N-1)\times S^{N-1}$.

Finally,  it remains to check that this map is well behaved 
with respect to the measures $\mu_{N-1}$, $\nu_N$ and $\mu_N$.
Given a function $f$ on $O(N)$, we may compute the average of $f$ with respect
to $\mu_N$ in two stages as follows:
First compute the conditional expectation $g(\vec v)$
where
$$g(\vec v) = {\rm E}\{ f(G)\ |\ \pi_N (G)= \vec v\}\ .$$
Then
$$\int_{O(N)}f(G){\rm d}\mu_N(G) = \int_{S^{N-1}}g(\vec v){\rm d}\nu_N(\vec v)\ .$$
Also, it is clear that we can compute the conditional expectation 
${\rm E}\{ f(G)\ |\ \pi_N(G) = \vec v \}$
by averaging over orbits generated by left multiplication of $G$ by elements 
$G'$ belonging to the subgroup
of $O(N)$ consisting of orthonormal matrices whose final row is ${\bf e}_N$. 
This subgroup is just a copy of
$O(N-1)$, and so
$$g(\pi_N(G)) = \int_{O(N-1)}f(G'G){\rm d}\mu_{N-1}(G')\ .$$
It follows directly from this that
$$\int_{O(N)}f(G){\rm d}\mu_N(G) =
\int_{S^{N-1}}\left[\int_{O(N-1)}f(\phi_N(G',\vec v)){\rm d}\mu_{N-1}(G')    
\right]{\rm d}\nu_N(\vec v)\ .$$
We define $\phi_j$ for $j=1,2,\dots,N-1$ in terms of $\phi_N$
and the pair permutations exchanging $j$ and $N$, in the natural way. Clearly the analogs
of the results
just derived for $\phi_N$ hold for each $\phi_j$ as well. Thus, with these definitions,
the third feature required of a
Kac system is verified.

To complete the verification that the $O(N)$ Kac walk can be made into a Kac system, 
we only need to verify that
$$\langle f,Qf\rangle_{{\cal H}_N} = {1\over N}\sum_{j=1}^N\int_{Y_N}\left(
\langle f_{j,\vec v},Qf_{j,\vec v}\rangle_{{\cal H}_{N-1}}
\right){\rm d}\nu_{N}(\vec v)\ $$
where for each $j$ and each $y\in Y_N = S^{N-1}$,
$$ f_{j,\vec v}(\cdot) = f(\phi_j(\cdot,\vec v))\ .$$
This is clear, given the definition and computations just above.

We are not yet ready to apply Theorems 2.1 and 2.2 to this Kac system, since we must 
modify the definition of $\lambda_N$. We define
$$\lambda_N = \sup\{
\langle f,Qf\rangle_{{\cal H}_N}\ |\ \|f\|_{{\cal H}_N}=1\ ,\ \langle f,1\rangle_{{\cal H}_N} = 0\ ,\  
\langle f,{\rm det}\rangle_{{\cal H}_N} = 0\ \}\ ,$$
which differs from \eqv(lamdef) due to the restriction that $f$ be orthogonal to the determinant.

As a consequence, a {\it  modification of the operators $P_j$ is also required}. In the definition \eqv(pjdef) used in examples
with a single ergodic component, we averaged over all of $X_{N-1}$, or put differently, conditioned
only on $Y_N$. Now we will condition on $Y_N$ and the ergodic component.
Given $f$ in ${\cal H}_N$, and $j = 1,2,\dots,N$, define $g_+$ and $g_-$ on $Y_N = S^{N-1}$ as follows:
$$g_+(\vec v) = 2\int_{O(N-1)_+}f_{j,\vec v}(\tilde G){\rm d}\mu_{N-1}(\tilde G)$$
and
$$g_-(\vec v) = 2\int_{O(N-1)_-}f_{j,\vec v}(\tilde G){\rm d}\mu_{N-1}(\tilde G) \ .$$
The definitions are such
that
$$g_\pm(\vec v) = {\rm E}\{f \ |\ \pi_j = \vec v\quad{\rm and} \quad {\rm det} = \pm 1\ \}\ .$$
The factors of $2$ are because $O(N-1)_\pm$ each accounts for exactly half of $O(N)$
by volume.
We now define
$$P_jf(G) = \cases{
g_+(\pi_j(G)) &if ${\rm det}(G) = +1$\cr
g_-(\pi_j(G)) &if ${\rm det}(G) = -1 \ .$\cr}
$$
Since $P_jf$ depends on $G$ only through $\vec v = \pi_j(G)$, it is again convenient to abuse 
notation by writing $P_jf(\vec v)$. 

The point of the definitions is the following: Note that  $f$ in
${\cal H}_N$ satisfies  both $\langle f,1\rangle_{{\cal H}_N} = 0$ and 
$\langle f,{\rm det}\rangle_{{\cal H}_N} = 0$ in case it satisfies both
$$\int_{O(N)_+}f( G){\rm d}\mu_{N}(G) = 0\qquad{\rm and}\qquad 
\int_{O(N)_-}f( G){\rm d}\mu_{N}(G) = 0\ .$$
But in this case, by the definition of $P_j$,
$$\int_{O(N-1)_+}\left(f_{j,\vec v}(\tilde  G)- P_jf(\vec v)\right){\rm d}\mu_{N-1}(\tilde G) = 0$$
and
$$\int_{O(N-1)_-}\left(f_{j,\vec v}(\tilde  G)- P_jf(\vec v)\right){\rm d}\mu_{N-1}(\tilde G) = 0\ ,$$
for almost every $\vec v$. Hence $\left(f_{j,\vec v}(\tilde  G)- P_jf(\vec v)\right)$ is, for almost every $\vec v$,
orthogonal to both $1$ and ${\rm det}$ on $O(N-1)$. This is the key requirement for the proof of Theorem 2.2
to hold with the modified definition of $\lambda_N$. We leave to the reader the easy verification of this.
Of course the definitions of the quantities $\mu_N$ in \eqv(mudef), 
and $\kappa_N$
in \eqv(kadef)  have to be modified in the same way as 
was the definition of $\lambda_N$. With these modifications made, the analog of Theorem 2.1 holds as well,
and again, as the proof is essentially the same, the verification of this is left to the reader.
{\it We summarize this by saying that the presence of more than one ergodic component can be taken into account
within the framework of ideas described in section 2 by conditioning not just on the single particle 
space, but on the ergodic components as well.} The present model is a case in point, which we choose to
leave as an example rather than attempting a general formulation.

With these results in hand, our task is to compute the spectrum of the 
$K$ operator for this system, which is still defined through \eqv(kdef). This is what is used
in the proof of Theorem 2.1, though \eqv(kker) no longer holds due to our modification
of the definition of $P_j$. 

However, it is clear from the
definition of $K$ in terms of conditional expectations, through \eqv(BQ2y), that
$$Kg(\vec v) = \int_{W\cdot \vec v=0}g(W){\rm d}\nu_{N-2}(W)\ ,\EQ(SN4)$$
where the integral on the right is computed with respect to the uniform 
probability measure on $S^{N-2}$, identified
with the subset of $S^{N-1}$ consisting of those unit vectors $W$ for which $W\cdot \vec v =0$,
 as indicated in the
limits of integration. That is, the value of $Kg(\vec v)$ is just the ``equatorial average'' of
 the values of $g$
around the ``equator'' with respect to a pole running along $\vec v$.

It is immediately clear that $K$ preserves the space of polynomials of 
any fixed degree $d$, and
hence the eigenfunctions of $K$ are the spherical harmonics on $S^{N-1}$.

The zonal spherical harmonics of degree $d$ are those that depend on 
$W$ only through ${\bf e}\cdot W$
for some fixed unit vector ${\bf e}$. Let $z_{d,\bf e}(W)$ denote the corresponding zonal
spherical harmonic, and let
$p_{d}(x)$ be the polynomial of degree $d$ so that the zonal spherical harmonic  
$z_{d,\bf e}(W)$ on
$S^{N-1}$ is given by
$$z_{d,\bf e}(W) = p_{d}({\bf e}\cdot W)\ .$$
The normalization is fixed so that the {\it reproducing kernel} property holds:
$$\int_{S^{N-1}}h(W)z_{d,\bf e}(W){\rm d}\nu_{N-1}(W) = h({\bf e})$$
for any spherical harmonic $h$ of degree $d$ on $S^{N-1}$. This means that
$$\|z_{d,\bf e}\|_2^2 = p_{d}({\bf e}\cdot {\bf e}) =p_d(1)\ .$$

Now fix $\vec v$, and let $h(W)$ be any spherical harmonic of degree $d$ that is orthogonal to
$z_{d,\vec v}$. Then
$$\int_{O(N-1)} h(\tilde G W){\rm d}\mu_{N-1}(\tilde G) = 0\Eq(SN6)$$
where $\tilde G$ runs over those rotations of $\R^N$ that fix $\vec v$. This is because the 
left hand side is a spherical harmonic
of degree $d$ that  depends on $W$ only through $\vec v\cdot W$. This means 
that it is a multiple of $z_{d,\vec v}$.
However, since $h$ was orthogonal to $z_{d,\vec v}$, so is the average, and hence
the identity  \eqv(SN6) is established. But comparing \eqv(SN4) and \eqv(SN6), we see that
$$Kg(\vec v) = \int_{O(N-1)} g(\tilde G W_0){\rm d}\mu_{N-1}(\tilde G)$$
for any $W_0$ with $W_0\cdot \vec v=0$. Hence, under our assumptions on $h$,  $Kh(\vec v) = 0$.

Now let $g$ be any spherical harmonic of degree $d$. Let $P_{\vec v}g$ be defined by
$$P_{\vec v} g(W) = {1\over p_d(1)}\left(
\int_{S^{N-1}}g(W')z_{d,\vec v}(W'){\rm d}\nu_{N-1}(W'))\right)z_{d,\vec v}(W) \Eq(SN8)$$
which is simply the orthogonal projection of $g$ onto the span of $z_{d,\vec v}$. 
Evidently $g- P_{\vec v}g$ is a
spherical harmonic of degree $d$ that is orthogonal to $z_{d,\vec v}$, and hence by the above,
$K(g- P\vec vg)(\vec v) = 0$. It follows that
$$Kg(\vec v) = K(P_{\vec v}g)(\vec v)\ .$$
But the right hand side is easy to compute as clearly $Kz_{d,\vec v} = p_d(0)$. 
Now by \eqv(SN8) and
the reproducing kernel property, we have
$$Kg(\vec v) = {p_d(0)\over p_d(1)}g(\vec v)\ .$$

Now it is possible to compute the ratios $p_d(0)/p_d(1)$ using generation functions,
though it would not be so clear from this
which value of $d$ gives the largest ratio.  However, none of this is necessary: If we fix
 any direction
unit vector $\vec e$, and take any function $\phi$ on $[-1,1]$, we have
$$K(\phi(\vec v\cdot \vec e)) = \left(K_{\rm Kac}\phi\right)(\vec v\cdot \vec e)$$
where $K_{\rm Kac}$ is the $K$ operator for the original Kac model, whose spectrum we have
computed in Theorem 3.1.
Hence the eigenvalues of $K$ and $K_{\rm Kac}$ coincide and are provided by Theorem 3.1

This solves the eigenvalue problem for $K$.

\medskip
\noindent{\bf Theorem 6.1} {\it Every spherical harmonic $\phi$ on $S^{N-1}$ of degree $d$, 
considered as a function $\tilde \phi$ on $O(N)$ through $\tilde\phi(G) = \phi(G\vec w)$ 
for any fixed vector $\vec w$ in $S^{N-1}$,
is an eigenfunction of
$K$, and the corresponding eigenvalue is $p_d(0)/p_d(1)$. These eigenvalues are the exactly the eigenvalues
of the $K$ operator for the original Kac model that are given in Theorem 3.1}
\medskip

Therefore, \eqv(goodform) and \eqv(finprod) hold just as in Section 3, and we have once more that
$$\Delta_N \ge {1\over 4}{N+2\over N-1}\Delta_2\ .$$

It remains to calculate $\Delta_2$. But  $O(2)$ is just two copies of $S^1$, and the same Fourier analysis
argument described around  \eqv(AQ2p)
once more gives us
$$\lambda_2 = \sup_{k\ne 0}
\left\{\int_{-\pi}^{\pi}\rho(\theta)\cos(k\theta){\rm d}\theta\right\}$$
for the second largest eigenvalue of $Q$ when $N= 2$. 
\medskip
The generalization of Maslin's result to nonuniform $\rho$ now follows
immediately from what has been said above.

Altogether, we have proved:

\medskip
\noindent{\bf Theorem 6.1} {\it The spectral gap for the Kac walk on $SO(N)$ with the transition operator $Q$
given by \eqv(SG6) coincides exactly, for each $N\ge 2$, with the spectral gap for the transition operator $Q_{\rm Kac}$
of the corresponding Kac walk on $S^{N-1}$, as given in \eqv(A6a) with the same density $\rho$. When $\rho$
is uniform, the multiplicity of the corresponding eigenvalue equals the dimension of the space of 
fourth degree spherical harmonics on $S^{N-1}$.}
\medskip

\bigskip

\chap {7:  Analysis of Maximizers for Nonuniform $\rho(\theta)$ }7

\bigskip

We return to the  Kac model on $S^{N-1}$  with a non uniform density $\rho(\theta)$,
so that $Q$ is given by \eqv(A6a), 
and we examine the circumstances under which the quartic function $f_N$ given in \eqv(evform1)
is an optimizer for \eqv(lamndef). Because of the very close relation of the Kac walk on
$SO(N)$ to this model, as described in Section 6, 
our  analysis is readily adapted to that model as well, 
though we shall not carry out the adaptation here.

According to Lemma 3.2 
$f_N$  is  an eigenfunction of $Q$ with the eigenvalue
$$
1-{2 \gamma (N+2) \over N(N-1)}\Eq(NUA1)
$$
where 
$$
\gamma = {1 \over 4} \int (1- \cos(4\theta)) \rho(\theta) {\rm d} 
\theta  \ .\Eq(NUA2)
$$
We therefore define $\Gamma_N$ by
$$\Gamma_N = {2 \gamma (N+2) \over (N-1)} = 
N\left(1 - {\langle f_N,Qf_N\rangle\over \|f_N\|^2}\right)\ ,\Eq(NU1)$$
so that if $f_N$ happens to be a maximizer for \eqv(lamndef), then $\Gamma_N = \Delta_N$,
but otherwise $\Delta_N < \Gamma_N  $. That is,
$$\Delta_N \le  \Gamma_N\ ,\Eq(NU2a)$$
and there is equality in \eqv(NU2a) if and only if $f_N$ is a maximizer for \eqv(lamndef).

Now the operator $Q$ commutes with permutations so the permutation invariant functions $f$
constitute an invariant subspace ${\cal S}$  of $\hh$, and clearly this subspace includes $f_N$.
It is the symmetric densities $f$ that are relevant in Kac's limit theorem relating the Master
equation \eqv(kacmast) and the Kac equation \eqv(kacequ). We will therefore restrict our attention
to this subspace, where the result is especially striking, and which is 
physically the most significant. We therefore define $\tilde \lambda_N$ to be the second largest 
eigenvalue of $Q$ restricted to ${\cal S}$:
$$\tilde \lambda_N = \sup\left\{{1\over N}\langle f, Q f\rangle\ \biggl|\ f\in {\cal S}\ , 
\ \|f\|_2=1\ ,\langle f, 1\rangle=0\ \right\}\ ,
\Eq(lam2def)$$
and we define $\tilde \Delta_N = N(1- \tilde\lambda_N)$. Clearly
$$\tilde \Delta_N \le  \Gamma_N\ ,\Eq(NU2)$$
and there is equality in \eqv(NU2) if and only if $f_N$ is a maximizer for \eqv(lam2def).

Taking the symmetry constraints into account it is easy to compute $\tilde \Delta_2$
using \eqv(AQ2p), with the result that
$$
\tilde\Delta_2 = 2 \min_{k \geq 1} \int(1-\cos(k \theta)) \rho(\theta)
{\rm d} \theta \ .\Eq(NU6)
$$
We see from \eqv(NUA1) and \eqv(NUA2) that if the supremum in \eqv(NU6) occurs at  $k=4$,
then 
$$\tilde \Delta_2 = \Gamma_2\ .\Eq(NU10)$$
It is easy to see that  $\tilde \Delta_N$ and $\tilde \Delta_{N-1}$
are still related by the inequality proved in Theorem 1.1 for $\Delta_N$ and $\Delta_{N-1}$:

$$\tilde \Delta_N \ge (1 - \kappa_N)\tilde \Delta_{N-1}\ .\Eq(NU12) $$
We also know from Theorem 1.3 and the definition of $\Gamma_N$ in terms of $f_N$ that $\Gamma_N$
solves this same recursion relation:
$$ \Gamma_N = (1 - \kappa_N) \Gamma_{N-1}\ .\Eq(NU13) $$

Notice that  \eqv(NU10), \eqv(NU12) and \eqv(NU13) together imply that 
$\tilde \Delta_N \ge \Gamma_N$
for all $N\ge 2$. But from \eqv(NU2) we have  $\tilde \Delta_N \le \Gamma_N$ for all $N\ge 2$. 
Hence $\tilde \Delta_N = \Gamma_N$ and $f_N$ is a maximizer. 

We see that if $f_2$
is the maximizer for $N=2$, the $f_N$ is a maximizer for all $N$.
So far we are simply translating old results into the symmetric case, 
but we have relied less on explicit calculation 
in order to bring out the following point: Suppose we had {\it any} sequence of admissible functions 
$g_N$ for the variational problem \eqv(lam2def),
and we defined $\Gamma_N$ by
$$\Gamma_N = N(1 - \langle g_N,Qg_N\rangle)\ .\Eq(NUa45)$$
Then if it happened that the $\Gamma_N$ so defined satsified the recurrence relation 
\eqv(NU13), and also satsified \eqv(NU10), it would follow by simple comparison that 
$\tilde \Delta_N = \Gamma_N$ for all $N\ge 2$ and $g_N$ would be a maximizer for \eqv(lam2def). 
{\it All that was 
required of $g_N$ is that \eqv(NUa45) leads to a
solution of \eqv(NU13), and that  \eqv(NU10) holds.}

The following simple observation leads to further progress: Suppose that the minimum in \eqv(NU6)
does not occur at $k=4$, and so  \eqv(NU10) is false. But suppose that for some $N_0$,
$\tilde \Delta_{N_0} = \Gamma_{N_0}$. That is, suppose that for $N= N_0$, $f_N$ is a maximizer for
\eqv(lam2def). {\it Then $f_N$ is a maximizer for \eqv(lam2def) for all $N\ge N_0$, so that
$\tilde \Delta_N = \Gamma_N$ for all $N\ge N_0$ and
therefore
$$\lim_{N\to\infty}\tilde \Delta_N = \lim_{N\to\infty}\Gamma_N = 2\gamma\ ,\Eq(NU56)$$
which is the result we would have gotten if the maximizer for $N=2$ had been quartic.
Therefore, either \eqv(NU56) holds, or else $f_N$ is never a maximizer for \eqv(lam2def) for
any $N$}

Now we know that $f_N$ spans the second eigenspace of $P$ corresponding to 
its second largest eigenvalue, where $P$ is the operator whose second largest eigenvalue $\mu_N$
is the key to the recursion in Theorem 2.2. (Recall that $P$ does not depend on $\rho$).
{\it If for each $N$, the true maximizer $h_N$ for
\eqv(lam2def) is orthogonal to $f_N$, which is the case whenever
$f_N$ is never a maximizer for \eqv(lam2def),
then we can replace $\mu_N$ in Theorem 2.2 by a smaller number
$\tilde \mu_N$, and hence can replace $\kappa_N$ in \eqv(NU12) by a smaller number $\tilde\kappa_N$.}
As we shall see, it turns out that this strictly smaller number $\tilde\kappa_N$ is
$$
\tilde\kappa_N = \alpha_8(N) = {105 \over (N+5)(N+3)(N+1)(N-1)} \ ,
\EQ(NU105)$$
where $\alpha(8)$ is an eigenvalue of the $K$ operator as described in Theorem 3.1, while 
$$\kappa_N = \alpha(4) = {3\over (N+1)(N-1)}\ .$$

In summary,  we have two things working for us:
\medskip
\noindent{\it Either \eqv(NU12) holds with a  $\kappa_N$ replaced by a strictly smaller number
$\tilde\kappa_N$ for all $N$, or else there is an $N_0$ so that $\tilde\Delta_N = \Gamma_N$ 
for all $N\ge N_0$. }
\medskip
Now when $\kappa_N $ is replaced by $\tilde\kappa_N$ in \eqv(NU12), 
it leads to a {\it much better} lower bound for $\tilde \Delta_N$. But this improved lower bound
cannot violate \eqv(NU2). If it does, it can only mean that the second alternative holds and not the first.

This argument leads to the following result:

\bigskip
{\bf Theorem 7.1: (Conditions for $f_N$ to Maximize for Large $N$)}\
{\it Assume that 
$$
\tilde \Delta_2 > 
0.45 \Gamma_2\ .\Eq(NU106)
$$
then for all $N$ sufficiently large, 
$$
\tilde \Delta_N = \Gamma_N \EQ(NU107)
$$
and $f_N$ is the corresponding eigenfunction.
} 

\medskip
\noindent{\bf Proof:}  All of the key ideas have been explained above, 
and it only remains to check the details.
As we have seen, if \eqv(NU107) does not hold for all sufficiently large $N$, 
then $f_N$ is orthogonal to the true gap eigenfunction
$h_N$ for all $N$, since Q is self adjoint and $f_N$ is always an eigenfunction.

This means that $h_N$
is orthogonal to the constant function {\it and} to the function $f_N$.
Now we repeat the induction argument in the proof of Theorem 2.2 once more 
but for the constraint that $h_N$ is orthogonal to both $1$ and $f_N$.
Under  these 
new conditions, we obtain the recursion
$$\tilde \lambda_N \le \left( \tilde \lambda_{N-1} + 
(1 - \tilde \lambda_{N-1})\tilde\mu_N\right)\ $$
in place of \eqv(basrec), where $\tilde \mu_N$ is given by
\eqv(mudef), except that now we require $f$ to be orthogonal to both $1$ and $f_N$.
Again, Theorem 2.1 shows that $\tilde \mu_N$ can be computed in terms of the eigenvalues of $K$,
with the result that it is $\alpha(8)$ that is now relevant, not $\alpha(4)$, due to the new
constraint. (Recall that $\alpha_6(N)$ is negative, and so is irrelevant.)
Hence
$$
\tilde \Delta_N \geq \tilde \Delta_{N-1}(1- \alpha_8(N)) \ , \Eq(NU108)
$$
The inequality \eqv(NU108) can be solved recursively to yield
$$
\tilde \Delta_N \geq 90 {\Gamma (N-1) \Gamma (N+7)\Gamma (N+3+i\sqrt 6)
\Gamma (N+3-i \sqrt 6) )
\over \Gamma(N) \Gamma(5+i\sqrt 6) \Gamma (5-i\sqrt 6)
\Gamma (N+6) \Gamma(N+4) \Gamma (N+2)} \tilde \Delta_2 \ . \Eq(NU90)
$$

With the help of the relation
$$
\Gamma(z)\Gamma(1-z) = {\pi \over \sin(\pi z)}
$$
 the limit as $N \to \infty$ of \eqv(NU90) can be computed and yields
$$
\liminf_{N \to \infty} \tilde \Delta_N \geq {3 \over 770}{\sinh(\sqrt 6 \pi) \over \sqrt 6 \pi} \tilde\Delta_2
=: L \tilde \Delta_2\ .
$$

This quantity has to be compared to $2 \gamma$ and this shows that whenever
$2\gamma < L \tilde \Delta_2$ (which is easily shown to be implied by \eqv(NU106))
there exists some finite $N$ beyond which
$f_N$ is the gap eigenfunction. \eop

\bigskip
\centerline{\bf References}
\vskip.3truecm

\item{[\rtag{A}]} Askey, R., {\it Orthogonal Polynomials and special Functions},
SIAM Regional Conferecne Series in Applied Mathematics, {\bf 21},
SIAM, Philadelphia, 1975

\vskip .3truecm

\item{[\rtag{TC}]} Carleman, T., Sur la solution de l'\'equation 
int\'egrodiff\'erentielle de Boltzmann, Acta Math., {\bf 60}, 91-146, 1933

\vskip .3truecm
\item{[\rtag{CCL}]} Carlen, E., Carvalho, M. and Loss, M.,
{\it Many--Body Aspects of Approach to Equilibrium}, in
{\it Journ\'ees \'Equations aux d\'eriv\'ees partielles}
N. Depauw et al eds., Nantes, June  2000.

\vskip.3truecm

\item{[\rtag{CGT}]} Carlen, E., Gabetta, E. and Toscani, G., {\it Propagation of
Smoothness and the Rate of Exponential Convergence to Equilibrium for a 
Spatially Homogeneous Maxwellian Gas}, Commun. Math. Phys. {\bf 205}, 521--546,
1999.

\vskip.3truecm

\item{[\rtag{DSC}]} Diaconis, P. and Saloff--Coste, L., {\it Bounds for Kac's
Master equation}, Commun. Math. Phys. {\bf 209}, 729--755, 2000.

\vskip.3truecm

\item{[\rtag{DS}]} Diaconis, P. and Shahshahani, M., {\it Generating a 
random permutation with random transpositions}, 
Z. Wahrsch. Verw. Gebiete {\bf 57}, 159--179, 1981.

\vskip.3truecm

\item{[\rtag{G1}]} Gruenbaum, F. A., {\it Propagation of chaos for the Boltzmann
equation}, Arch. Rational. Mech. Anal. {\bf 42}, 323--345, 1971.

\vskip.3truecm

\item{[\rtag{G2}]} Gruenbaum, F. A., {\it Linearization for the Boltzmann
equation}, Trans. Amer. Math. Soc.  {\bf 165}, 425--449, 1972.

\vskip.3truecm

\item{[\rtag{J}]} Janvresse, E., {\it Spectral Gap for Kac's model of 
Boltzmann Equation}, To Appear in Annals. of Prob., 2001.

\vskip.3truecm

\item{[\rtag{K}]} Kac, M., {\it Foundations of kinetic theory}, Proc. 3rd Berkeley
symp. Math. Stat. Prob., J. Neyman, ed. Univ. of California, vol 3,
pp. 171--197, 1956.

\vskip.3truecm

\item{[\rtag{Ko}]} Koornwinder, T.H., {\it The addition fromula for Jacobi polynomials. I,
summary of results}, Indag. Math. {\bf 34}, 188--191, 1972.

\vskip.3truecm

\item{[\rtag{M}]} McKean, H., {\it Speed of approach to equilibrium for Kac's caricature
of a Maxwellian gas}, Arch. Rational Mech. Anal. {\bf 21}, 343--367, 1966.

\vskip.3truecm

\item{[\rtag{Y1}]} Yau H.T., E., {\it The Logarithmic Sobolev Inequality for
Generalized Simple Exclusion Processes}, Probab. Theory and Related Fields. {\bf 109}, 507--538,
1997.

\vskip.3truecm

\item{[\rtag{Y2}]} Yau H.T., E., {\it The Logarithmic Sobolev Inequality for
Lattice Gasses with Mixing Conditions}, Commun. Math. Phys. {\bf 181}, 367--408,
1996.

\end